\documentclass[11pt,a4paper]{article}
\usepackage[utf8]{inputenc}
\usepackage{graphicx,verbatim,array,multicol,courier}
\usepackage{amsmath,amssymb,amsthm,mathrsfs, mathtools}
\usepackage{color, graphics, graphicx, siunitx}
\usepackage{amsfonts, dsfont, bm}
\usepackage{tabularx}
\usepackage{natbib} 
\usepackage{geometry}
\usepackage{enumitem,float}
\usepackage{lscape}
\usepackage[pdftex,bookmarks,colorlinks]{hyperref}
\usepackage[dvipsnames]{xcolor}
\usepackage{siunitx}
\usepackage[labelformat=simple]{subfig}
\usepackage[linesnumbered,ruled,vlined]{algorithm2e}
\usepackage{booktabs}

\newlist{inparaenum}{enumerate}{2}
\setlist[inparaenum,1]{label=(\alph*)}
\setlist[inparaenum,2]{label=(\roman{inparaenumi}\emph{\alph*})}

\makeatletter
\def\adl@drawiv#1#2#3{%
        \hskip.5\tabcolsep
        \xleaders#3{#2.5\@tempdimb #1{1}#2.5\@tempdimb}%
                #2\z@ plus1fil minus1fil\relax
        \hskip.5\tabcolsep}
\newcommand{\cdashlinelr}[1]{%
  \noalign{\vskip\aboverulesep
           \global\let\@dashdrawstore\adl@draw
           \global\let\adl@draw\adl@drawiv}
  \cdashline{#1}
  \noalign{\global\let\adl@draw\@dashdrawstore
           \vskip\belowrulesep}}
\makeatother

\numberwithin{equation}{section}
\theoremstyle{definition}
\newtheorem{defi}{Definition}[section]
\newtheorem{cond}[defi]{Condition}
\theoremstyle{plain}

\newtheorem{prop}[defi]{Proposition}

\newtheorem{cor}[defi]{Corollary}

\theoremstyle{remark}
\newtheorem{rem}[defi]{Remark}

\theoremstyle{example}
\newtheorem{ex}[defi]{Example}

\newcommand{\diff}{\mathrm{d}}

\definecolor{navy}{rgb}{0,0,0.502}
\definecolor{brown}{rgb}{0.59, 0.29, 0.0}

\hypersetup{colorlinks,%
	citecolor=blue,%
	filecolor=green,%
	linkcolor=red,%
	urlcolor=violet,%
}

\newcommand{\hell}{{\mathscr{H}}}

\newcommand{\Real}{\mathbb{R}}

\newcommand{\Prob}{\mathbb{P}}

\newcommand{\bftheta}{{\boldsymbol{\theta}}}
\newcommand{\bfvartheta}{{\boldsymbol{\vartheta}}}

\newcommand{\bfSigma}{{\boldsymbol{\Sigma}}}
\newcommand{\bfbeta}{{\boldsymbol{\beta}}}
\newcommand{\bfvarepsilon}{{\boldsymbol{\varepsilon}}}

\newcommand{\bfX}{{\boldsymbol{X}}}
\newcommand{\bfY}{{\boldsymbol{Y}}}

\newcommand{\bfy}{{\boldsymbol{y}}}

\title{Statistical Prediction of Peaks Over a Threshold}
\author{S. A. Padoan\\
	Department of Decision Sciences, Bocconi University, Italy\\
	and\\ 
	S. Rizzelli\\
	Department of Statistical Sciences, University of Padova, Italy
}

\begin{document}

\maketitle
\begin{abstract}
In many applied fields, the prediction of more severe events than those already recorded is crucial for safeguarding against potential future calamities. What-if analyses, which evaluate hypothetical scenarios up to the worst-case event, play a key role in assessing the potential impacts of extreme events and guiding the development of effective safety policies.
This problem can be analyzed using extreme value theory. We employ the well-established peaks-over-threshold method and describe a comprehensive toolkit to address forecasting needs. We examine an \lq\lq out-of-sample" variable and focus on its conditional probability of exceeding a high threshold, representing the predictive distribution of future extreme peaks. We demonstrate that the generalized Pareto approximation of the corresponding predictive density can be remarkably accurate. We then introduce frequentist methods and a Bayesian approach for estimating this predictive density, enabling the derivation of informative predictive intervals. By leveraging threshold stability, we illustrate how predictions can be reliably extended deep into the tail of the unknown data distribution. We establish the asymptotic accuracy of the proposed estimators and, more importantly, prove that the resulting predictive inference is 
asymptotically valid. Forecasters satisfying the tail-equivalence property allow to recover widely used risk measures for risk assessment through point forecasts. This insight lays the groundwork for a new perspective that integrates risk assessment into the statistical predictive toolbox. Finally, we extend the prediction framework to the case of linear time series. We apply the proposed predictive tools to two real-world datasets: summer peak temperatures recorded in Milan, Italy, over the past 30 years, and daily negative log-returns of the Dow Jones Industrial Average observed over 30 years.
\end{abstract}


%
\section{Introduction}\label{sec:intro}
%
%
\subsection{From risk assessment to statistical prediction of extremes}

In fields such as environmental sciences, finance, insurance, and other applied domains, there is a strong interest in assessing the risk of future events. In risk assessment, the standard approach to evaluating the risk of a specific event involves estimating a risk measure, which is statistical property of the distribution $F$ of a random variable $X$, representing a quantity of interest. Risk measures are indexed by a ``level''  $\tau$, which defines the frequency and severity of certain events \citep[e.g.,][]{artzner1999coherent}.
In economics and finance, commonly used risk measures to quantify potential losses include the Value at Risk (VaR, or quantile) and the Expected Shortfall (ES, or conditional tail expectation), see, e.g., \cite{ivete2013, marie2015, marie2018}. In the case of independent data, a widely accepted interpretation of the $\tau$-quantile $Q(\tau)$ corresponding to a distribution $F$ is given by setting $\tau=1-1/T $, where $T$ is the {\it return period}. In this context, $Q(\tau)$ is interpreted as a  {\it return level}, which is the level that is expected to be reached or exceeded once every $T$ time units (e.g. days, months, years; see \citealp{rootzen2013design} for broad discussion).

Extreme Value Theory (EVT) provides the probabilistic and statistical framework to extrapolate
tail events into the far tail of an unknown data distribution, helping to assess the plausibility and magnitude of future episodes that may  exceed  those previously observed
 \citep[e.g.][]{leadbetter1983extremes, coles2001, dehaan+f06, de2016statistics}.
The literature on tail risk assessment primarily focuses on estimating extreme risk measures. Among the most commonly used is the {\it extreme quantile} (or extreme VaR)---a quantile of the data distribution expected to exceed the largest observed value of the sample. Given a sample $\bfX_n=(X_1\ldots,X_n)$ of independent and identically distributed (iid) random variables ith distribution $F$ and  $\tau$-quantile $Q(\tau)$, the relevant scenario occurs when $\tau=\tau_n$ where $\tau\to 1$ and $n(1-\tau)\to \nu<1$, as the sample size  $n\to\infty$. In this extrapolation regime, we expected no exceedances above  such a quantile \citep[e.g.,][Ch. 4]{dehaan+f06}. 

Extrapolating risk measures involves estimating the magnitude of an event, with a tiny probability of being exceeded, whose evaluation being subject solely to estimation (epistemic) uncertainty. However, this approach fails to account for aleatoric uncertainty, which arises from the inherent randomness and unpredictability of future occurrences within the phenomenon under study \citep[e.g.,][]{BJARNADOTTIR2019271}. In this regard, forecasting provides a means to incorporate aleatoric uncertainty when making claims about the future. As highlighted more than forty years ago by \cite{dawid1984present}, forecasts should be expressed as probability distributions over future events, as probability is the most meaningful way to disclose uncertainty. From a statistical perspective, predictions are made by specifying predictive distributions \citep[][to name a few]{Aitchison_Dunsmore_1975, geisser1993predictive, BERNARDO2011263}. In general, given a sample $\bfX_n$ of independent or dependent variables representing past observations, the statistical goal is to infer the distribution $F^\star$ of an ``out-of-sample'' variable $X_{n+1}$, which represents future, unobserved yet events and may be either independent or dependent on $\bfX_n$. The estimated $F^\star$  is then used to forecast the occurrence and magnitude of future events. In parametric settings, a {\it plug-in} approach is to estimate the parameters of $F^\star$ using past data and then plug these estimates into the model to obtain an approximate predictive distribution. A more rigorous frequentist alternative is based on the predictive likelihood  \citep[e.g.,][]{hinkley1979predictive, butler1986predictive}, where future observations depend on past data through conditioning on a sufficient statistic. Unfortunately, in extreme value analysis, no sufficient statistics are available, limiting the applicability of this approach. In the Bayesian framework, a natural and widely used method for statistical prediction is based on the posterior predictive distribution, which is obtained by integrating out the model parameters  with respect to their posterior distribution. Pioneering works on the posterior predictive distribution in extreme value context have been  \cite{davison1986approximate, coles1996bayesian, smith1997predictive, smith1999, coles2003}. 

Forecasts must be as reliable as possible to be useful, particularly when it comes to extreme events, as in many applied fields there is a pressing need to predict future calamities—such as global financial crises or severe monetary losses due to extreme weather—to provide decision-makers with a clear understanding of the associated uncertainty. In this context, tailored methods that offer accuracy guarantees are essential. 
Within extreme value analysis, \cite{coles1996bayesian, coles2003} empirically demonstrated the superiority of Bayesian predictive inference over frequentist approach across various applications, while  \cite{smith1997predictive, smith1999} were the first  methodological
works where the performance of predictive distributions derived from the frequentist plug-in approach and the Bayesian method have been formally compared. However, these studies simplify the complexity of the problem by adopting the so-called {\it vanilla} approach. Specifically, when the data distribution is unknown, as is often the case in applications, EVT provides approximate extremal models for analyzing extreme events that would otherwise be difficult to study (e.g., \citealp[][Ch. 3-4]{coles2001}; \citealp[][Ch. 1]{dehaan+f06}). Because these models are only asymptotically valid, they are inherently misspecified from a statistical perspective, regardless of the sample size.
The vanilla approach erroneously assumes that extremes follow the extremal models exactly, leading to the naive conclusion that estimators of extreme model parameters are asymptotically unbiased. In practice, this misspecification complicates the asymptotic theory of estimators, as it requires controlling the rate at which the actual distribution of extreme values converges to the extremal model via {\it second-order conditions} and handling complex empirical processes (e.g., \citealp{drees04}, \citealp[][Ch. 2, 4]{dehaan+f06}, \citealp{dombry19}). Moreover, extreme value models are {\it irregular} in that their support depends on the sign of the shape parameter, further complicating estimation \citep[e.g.][]{smith1985maximum, b+s2017}. As a result, developing mathematically rigorous methods for estimating predictive distributions under extremal models remains a challenging task.
It is therefore not surprising that Smith’s seminal works did not receive the extensive follow-up they deserved. Nevertheless, recent advances have begun to shed light on the accuracy of predictive distributions within the Bayesian framework, see \citet{padoan24b, padoan24}, and \citet{dombry23}.

\subsection{Contributions and outline}

This paper provides an extensive and systematic study on the accuracy of statistical prediction for extreme events using the celebrated Peaks Over Threshold (POT) method \citep{davison1990}, which is the most popular approach for analyzing univariate tail events. We begin by establishing sufficient conditions under which the Generalized Pareto (GP) density provides an accurate approximation of density of $\Prob(X \leq x \mid X>t)$, for a threshold $t$ approaching the right end-point of $F$ and a large value $x>t$, using the Hellinger distance \citep[e.g.,][]{vdv2000}. Next, given a sample $\bfX_n$, we consider an``out-of-sample'' variable $X_{n+1}$ and focus on predicting future extreme events by studying $\Prob(X_{n+1}\leq x \mid X_{n+1}>Q(\tau), \bfX_n)$, i.e. the predictive distribution of future peaks exceeding a sufficiently large threshold $Q(\tau)$. We allow $Q(\tau)$ to increase at different rates with sample size, 
where fast (slow) threshold growth corresponds to $\nu$ approaching zero (infinity). In this way we can push the conditioning threshold in the very far tail of the data distribution, enabling for a comprehensive {\it What-if} analysis for risk assessment, allowing for the evaluation of hypothetical scenarios up to the worst-case events, thereby helping to understand the potential societal impacts of exceptional extreme events. In disaster risk analysis, incorporating rare events is essential, as excluding worst-case scenarios can lead to a significant underestimation of total risk. Leveraging the {\it threshold stability} property from EVT we introduce a class of GP approximations for the true, unknown predictive distribution and its density function, demonstrating that the latter achieves high accuracy in Hellinger distance. We then propose both frequentist and Bayesian methods for estimating the unknown predictive distribution and its density, proving that the estimated densities remain asymptotically close to the true densities in Hellinger distance, with probability tending to one. More importantly, when an extreme region is defined as a subset of the estimated peaks' support with a predictive probability level of $1-\alpha$, we establish that its coverage probability under the true peaks-generating distribution asymptotically reaches the nominal level $1-\alpha$.
Furthermore, forecasters (as a predictive distribution) satisfying the {\it tail-equivalence} property relative to the true predictive distribution, allow to recover  standard risk measures commonly discussed in the literature. We illustrate how to construct point forecasts from our frequentist and Bayesian predictive distributions that provide consistent estimators of well-known risk measures such as VaR and ES. These findings introduce a novel perspective: by embedding tail risk assessment within a statistical prediction framework, we achieve a significant advantage—predictive intervals not only cover specific point predictions of future peaks (which are informative about traditional risk measures) but also quantify prediction uncertainty. 
Finally, we describe how to perform one-step-ahead prediction of large peaks in linear time series and showcase the utility of our predictive tools through two real-world applications. We analyze extreme summer temperatures recorded in Milan, Italy, over the past thirty years, and the daily negative log-returns of the Dow Jones index over recent decades, demonstrating how to dynamically predict large peaks using point forecasts and predictive intervals.

The online supplementary material contains technical results and detailed proofs. Section \ref{sec:background} provides a brief background on EVT and introduces our first theoretical result. In Section \ref{sec:strong_results}, we analyze the GP approximation of the true predictive densities for different threshold magnitudes and assess their accuracy.  In Sections \ref{sec:freq} and \ref{sec:bayes}, we introduce estimators for predictive densities and establish their asymptotic accuracy. In Section \ref{sec:tail_risk_assessment}, we show how popular extreme risk measures can be consistently estimated using point forecasts derived from our predictive distributions. Section \ref{sec:time_series} extends our methodology to the case of linear time series. We apply the proposed methods in Section \ref{sec:real_analysiss}. Finally, Section \ref{sec:conclusion} concludes the paper with a discussion on future research directions.

%
\section{Background}\label{sec:background}
%
\subsection{Approximate distribution of extremes}

Let $X$ be a random variable following an unknown distribution $F$, whose right end-point is $x_{E}$. EVT provides an engine for assessing extreme events relying on the condition that $F$ belongs to the  {\it domain of attraction} of a Generalized Extreme Value (GEV) distribution—a weaker requirement than the restrictive assumption of knowing $F$ entirely \citep[][Ch. 1.2]{dehaan+f06}.
The domain of attraction condition simply states that  that for $n=1,2,\ldots$ if  
there are functions $a(n)>0$ and $b(n)$ such that $F^n(a(n)x+b(n))$ converges pointwise to a nondegenerate distribution as $n\to \infty$, then  such a limit must be indeed the GEV distribution $G_\gamma(x)=\exp(-(1+\gamma x)^{-1/\gamma})$, for all $x$ such that $1+\gamma x >0$, with $\gamma\in\Real$ \citep[e.g.,][Theorem 1.1.3]{dehaan+f06}.  The {\it extreme value index} $\gamma$ describes the weight of the upper tail of the distribution, i.e. if $\gamma>0, \gamma=0$ or $\gamma<0$ the distribution is {\it heavy-}, {\it light-} or {\it short-tailed}, respectively. Note that in \cite{coles2001} such a shape parameter is denoted by $\xi$.

For a threshold  $t<x_{E}$ and all $x>0$ let
\begin{equation}\label{eq:ConDist}
	F_t (x) = 
	\frac{F(x+t)-F(t)}{1-F(t)}.
\end{equation}
be the conditional distribution  of the random excess $X-t$ given that $X>t$.
The domain of attraction condition implies the following important result. If for all $t< x_E$, 
there is a scaling function $s(t)>0$ such that the pointwise convergence result $F_t(s(t)x)\to H_\gamma(x)$ as $t\to x_E$ holds, then $H_\gamma$ must be 
the standard GP distribution  \citep[][Theorem 1.2.5]{balkema1974residual, dehaan+f06}
\begin{equation}\label{eq:GP}
	%
	H_\gamma(x):=
	\begin{cases}
		1-\left(1+\gamma x\right)^{-1/\gamma},\quad x \in \mathcal{S}_\gamma,
		\\0, \hspace{8em}
		\text{otherwise},
	\end{cases}
\end{equation}
where $\mathcal{S}_\gamma=(0,\infty)$ if $\gamma\geq 0$ and $\mathcal{S}_\gamma=(0,-1/\gamma)$ otherwise.
A standard option is to select $s(t)= a(1/(1-F(t)))$, where $a(\cdot)$ is 
scaling function used in the block-maxima method \citep[][Ch. 1]{dehaan+f06}. Hereafter we stick to this choice.
The GP density function is
$$
h_\gamma(x)=\left(1+\gamma x \right)^{-(1/\gamma+1)},\quad x\in \mathcal{S}_\gamma.
$$ 

The above result is useful for statistical modeling because, even though $F$ is unknown,  by choosing a large threshold $t$  and a positive value $y$ (e.g. $y\equiv y_t=s(t)x$), then the conditional distribution $F_t(y)$ of the excess $X-t$, given that $X>t$, can be approximated by the two-parameters GP distribution $H_{\bftheta}(x):=H_\gamma(x/\sigma)$, with $\bftheta=(\gamma, \sigma)$ and where $\sigma>0$ is a scale parameter representative of $s(t)$. Its density is $h_{\bftheta}(x):=h_\gamma(x/\sigma)/\sigma$ for $x\in\mathcal{S}_\bftheta$, with $\mathcal{S}_\bftheta=(0,\infty)$ if $\gamma\geq 0$ and $\mathcal{S}_\bftheta=(0,-\sigma/\gamma)$ otherwise. Furthermore,
the random {\it peak} $X$, given that $X>t$, 
 provides a notion of extreme value with an intuitive  interpretation for applications, whose conditional distribution $F_t(y-t)$ can be approximated by $H_{\bftheta}(y-t)$ for
large enough $t$ and  $y$ (e.g. $y\equiv y_t=t+s(t)x$). 
These approximations are of practical relevance, providing the basis  for the extrapolation of extreme risk measures. They allow indeed 
to also approximate  the unconditional distribution for $y> t$ as follows
$$
F(y)\approx 1-(1-F(t))(1-H_{\bftheta}(y-t)), \quad t\to x_E.
$$
Furthermore, let $Q(\tau)=F^{\leftarrow}(\tau)$ be the $\tau$-quantile for any $\tau\in[0,1]$, where $F^{\leftarrow}(z)=\inf\{x:F(x)\geq z\}$ is the left-continuous inverse function of $F$. Let also
$$
E(\tau)=\frac{1}{1-\tau}\int_\tau^1Q(x) \diff x,
$$
be the popular risk measure ES.
Applying the above approximation to the equation $\tau=F(y)$ we have on one hand that the extreme VaR $Q(\tau)$ can be approximated by
\begin{equation}\label{eq:extreme_VaR}
	Q(\tau)\approx t + \sigma\frac{\left(\frac{1-\tau}{1-F(t)}\right)^{-\gamma}-1}{\gamma}, \quad \tau\to 1,
\end{equation}
and on the other hand that ES can be approximated as $\tau\to 1$ by 
\begin{equation}\label{eq:extreme_ES}
	E(\tau)\approx
	\begin{cases}
		(1-\gamma)^{-1} Q(\tau), & \text{ if } \gamma\geq 0,\\
		Q(\tau), & \text{ if } \gamma< 0,\\
	\end{cases}
\end{equation}
provided that $\gamma<1$. Accordingly, the estimation of such risk measures only requires in turn the estimation of $F(t)$, $\gamma$ and $\sigma$ from the data, once a large threshold $t$ is chosen. 
%

Another key contribution of EVT to risk assessment for events more extreme than those observed is the threshold stability property \citep[e.g.,][Ch. 1, 5]{falk2010}, which underpins the GP-based approximation.
Loosely speaking, a random variable $Y$ follows 
a threshold stable distribution if applying a thresholding operation to $Y$ preserves its distribution, up to a scaling factor. More formally, if $Y$ follows $H_\bftheta$ and, for $u\geq0$, we have $H_\bftheta(u)<1$ and $\sigma+\gamma u>0$, then 
the distribution of the random excess $Y-u$, given that $Y>u$, is 
$H_{\bfvartheta}$, with $\bfvartheta=(\gamma, \sigma+\gamma u)$. 
This property has significant practical implications: when analyzing the excess $X-t^\star$, given that $X>t^\star$ for some higher threshold $t^\star\geq t$, its distribution can still be approximated by a GP distribution of the form $H_{\gamma}(y/s(t^\star))$. However, the scale parameter is now updated to $s(t^\star)=s(t)+\gamma(t^\star-t)$.
This feature is particularly valuable in applications, as it allows for a progressive increase in the threshold while still relying on the GP distribution—albeit with an appropriately adjusted scaling parameter—to extrapolate extreme events further into the tail of the data distribution.

The following sections outline statistical prediction methods conditioned on the occurrence of tail events. 
In many applications, particularly in physics and climatology, it is reasonable to assume that the data distribution admits a density function. Therefore, we consider both distribution-based and density-based forecasting, establishing key results that ensure their reliability. We  demonstrate how these predictive tools can be effectively used to obtain consistent estimates of extreme risk measures. For clarity, we denote the density of $F_t(y)$ by $f_t$.

\subsection{Quality of the approximation}
In this section, we present a key probabilistic approximation that serves as the technical foundation for the statistical results that follow.
In this regard, we recall that the domain of attraction condition, also known as the {\it first-order condition}, that leads to the GP distribution in  \eqref{eq:GP} can be equivalently stated 
in terms of the the following convergence
\begin{equation}\label{eq:quant}
\lim_{u\to 0}\frac{Q(1-ux)-Q(1-u)}{a(1/u)}=\mathcal{Q}_\gamma(x):=\frac{x^{-\gamma}-1}{\gamma},
\end{equation}
for all $x>0$. 
Studying the speed with which $F_t$  converges pointwise to 
 $H_{\gamma}$ we can understand the goodness of the approximation that the Pareto distribution offers. To this purpose, one can use the following second-order condition.
\begin{cond}\label{cond:SO_basic}
%
There is a
function $A$, called rate (or second-order auxiliary) function such that $A(t')\to 0$ as $t'\to\infty$, $|A(t')|$ is regularly varying with index  $\rho\leq 0$, named second-order parameter,  and
$$
\lim_{u\to 0} \frac{1}{A(1/u)}\left(\frac{Q(1-ux)-Q(1-u)}{a(1/u)}-\mathcal{Q}_\gamma(x) \right)= \int_ 1^{1/x} v^{\gamma-1} \int_1^v r^{\rho-1} \diff r \diff v 
$$
see \citet[][Ch. 2]{dehaan+f06} for details.
\end{cond}
Stronger convergence forms than that used to obtain the formula \eqref{eq:GP} have been established under suitable conditions at the density level \citep[see e.g.][]{raoult03, padoan24}. 
The convergence result in Hellinger distance from \cite{padoan24} is particularly valuable for statistical applications, especially in the context of statistical prediction. This result is established under the following second-order von Mises-type condition, which specifies a particular rate function \citep[see][Condtion 4]{deHaan96, raoult03}. 
\begin{cond}\label{cond:SO}
The distribution function $F$ is twice differentiable, and 
Condition \ref{cond:SO_basic} holds with a rate function of the form
$$
A(t'):= \frac{Q^{(2)}(1-1/t')/t' -2 Q^{(1)}(1-1/t')}{Q^{(1)}(1-1/t')}+1-\gamma,
$$
where $Q^{(j)}(x):=(\partial^j /\partial x^j) Q(x)$, $j=1,2$.
\end{cond}
As shown in the supplementary material of \citet{padoan24b}, the majority of statistical models with a well-defined density satisfy this condition.

Let $l_t(x)= f_t( a(t')x) a(t')$ be the density of the conditional distribution $F_t(a(t')x)$, where we set $a(t')=s(F^{\leftarrow}(1-1/t'))$ with  $t'=1/(1-F(t))$.
Under Condition \ref{cond:SO}, \cite{padoan24} established that for $\gamma \geq 0$ the density $l_t$  converges to the density  $h_{\gamma}$  in Hellinger distance, denoted here as $\hell$, with 
speed of convergence  $|A(t')|$.
Here we extend that result  in order to cover the more general case that  $\gamma>-1/2$, which is typically considered when using  important inferential tools as the maximum likelihood method and Bayesian approach 
to analyse  peaks over threshold \citep{drees04, dehaan+f06, dombry23}. 
In particular, the GP family satisfies some regularity conditions useful for the likelihood-based inference, see \citet{dombry23}, and for its density we provide further the following approximation result.
\begin{prop}\label{theo:hellrate}
Assume Condition \ref{cond:SO} is satisfied with $\gamma >-1/2$. Let $\bftheta=(\gamma,\sigma)$ with $\sigma =a(t')$. Then, there exist constants $0<c<C<\infty$ and $t_0< x_E$ such that, for all $t \geq t_0$,
\begin{equation}\label{eq:hellrate}
c |A(t')| \leq \hell(l_t, h_\bftheta ) \leq C |A(t')|{.}
\end{equation}
\end{prop}
The following sections show the statistical utility of the above result, as it plays an important role in controlling bias in density estimation for the POT method,  and, in turn, it facilitates the assessment of the accuracy of the corresponding predictive distribution for tail events.

\section{Statistical prediction}\label{sec:strong_results}

In statistics, predicting future values is a fundamental task with wide-ranging applications. In this context, statistical prediction provides as a powerful tool for determining the entire predictive distribution (or its density) of future events, enabling the forecasting of their occurrence and severity. The predictive density, for instance, can be used to construct reliable predictive regions.
Prediction becomes particularly challenging when the focus is on extreme future events that exceed 
those previously observed. Here, we introduce a straightforward yet effective predictive approach, conditioned on tail events, leveraging models and statistical methods from EVT. To ensure their practical utility, predictions must be as accurate as possible. Given that our proposed tools are based on asymptotically justified models, we further investigate formal mathematical guarantees that validate their reliability.

Let $\bfX_n$ be a sample of independent and identically distributed (iid) random variables with a common unknown distribution $F_0$. For clarity and ease of notation, we omit the subscript “$0$” from true distributions, densities, parameters, etc., throughout the main text, except where its inclusion is strictly necessary, as detailed in the supplement. 
Next, we consider an \lq\lq out-of-sample" variable, $X_{n+1}$, representing future events. Since our focus is on extreme events, we examine the conditional distribution $F_{t}^\star(y):=\Prob(X_{n+1}\leq y \mid X_{n+1}>t, \bfX_n)$, which characterizes the distribution of future, yet-unobserved peaks, conditionally they exceed a danger threshold $t$. We refer to it as the predictive distribution of tail events.
EVT suggests the following procedure for selecting $t$. First, choose the number $k=k_n<n$ of peak variables from the sample to be used for estimating $F_{t}^\star$. This quantity, known as the {\it effective sample size}, should satisfy the conditions $k\to \infty$ as $n\to\infty$ and $k=o(n)$. Next, define the {\it intermediate level} $\tau_I=1-k/n$, where $k/n$ represents a small sample fraction. The value of $k$ should be suitably selected—small enough to ensure that the intermediate threshold $t_I=Q(\tau_I)$ is sufficiently large, yet not so small so that $t_I$ does not fall outside the data sample on average. According to Section \ref{sec:background},  $F_{t_I}^\star(y)$ and its density $f_{t_I}^\star(y)$ can be approximated by $H_\bftheta(y-t_I)$ and $h_{\bftheta}(y-t_I)$ (under  Condition \ref{cond:SO}), respectively, where $\sigma$ is representative of $a(t_I')$ with $t_I'=1/(1-F(t_I))$.

However, in many applications, ensuring safety requires a clear understanding of the risks associated with the extending of a danger threshold deep into the tail of the data distribution to explore the potential “worst-case scenario.” For this purpose, we introduce an {\it extreme level} $\tau_E=1-\nu/n$ for $\nu \equiv \nu_n \in (0,k ]$, allowing it to be closer to 1 than $\tau_I$. This, in turn, defines the extreme threshold $t_E=Q(\tau_E)$. 
To monitor potential hazards and assess associated risks, we introduce the parameter $\tau^\star\equiv \tau_n^\star=(1-\tau_E)/(1-\tau_I)$ and assume that $\lim_{n\to\infty} \tau^\star =\tau_0\in[0,1]$. When $\tau_0$ is close to $0$, then $t_E$ is much bigger than $t_I$, while $\tau_0$ close to $1$ indicates they are similar.
Exploiting the threshold stability property, the GP-based approximation of the predictive distribution $F_{t_E}^\star(y)$ and its density $f_{t_I}^\star(y)$ is given by  
\begin{align}
\label{eq:GP_predictive_dist}
H_{\bftheta,t_E}^\star(y)&=H_\bftheta
	\left(
	(y-t_I){\tau^\star}^{\gamma}-
	\sigma\frac{1-{\tau^\star}^{\gamma}}{\gamma}
	\right),\\
\label{eq:GP_predictive_dens}
h_{\bftheta, t_E}^\star(y)&=h_\bftheta
	\left(
	(y-t_I){\tau^\star}^{\gamma}-
	\sigma\frac{1-{\tau^\star}^{\gamma}}{\gamma}
	\right)
	\tau^\star{\color{red}.}
\end{align}
The following corollary, building on the result of Proposition \ref{theo:hellrate}, establishes that the densities $h_{\bftheta, t_E}^\star$ provide an accurate approximation of the true density $f^\star_{t_E}$ across a range of extreme levels $\tau_E\geq \tau_I$.
\begin{cor}\label{cor:newapprox}
Assume that Condition \ref{cond:SO} is satisfied with  $\gamma>-1/2$ and, when $\rho=\tau_0=0$,
that $|A|$ is nonincreasing near infinity. If
$A(n/k) w_{\gamma}(1/\tau^\star)\to0$ as $n\to\infty$, with
$$
w_{\gamma}(x) := \begin{cases}
			\log(x), \hspace{1.3em} \gamma>0,\\
			\log^2(x), \quad \gamma=0,\\
			x^{-\gamma}, \hspace{2.4em} \gamma<0,
		\end{cases} 
		$$
		where $x>0$, we have
		$
		\hell(h_{\bftheta, t_E}^\star, f_{t_E}^\star)=O\left( \sqrt{A(n/k)w_{\gamma}(1/\tau^\star))}\right).
		$
\end{cor}
Corollary \ref{cor:newapprox} plays a key role in demonstrating, in the following sections, that GP-based estimators of the predictive distribution and its density provide reliable inferential tools. Before describing specific estimation methods we discuss two important aspects. First, to make formulas \eqref{eq:GP_predictive_dist} and \eqref{eq:GP_predictive_dens} practically operational, we need to estimate the true intermediate threshold $t_I$. Approximating the unknown distribution $F$ with the empirical distribution $F_n$ has two key implications: (i) since $t_I$ is given by $F_n^{\leftarrow}(\tau_I)$, it can be approximated by the $(n-k)$th order statistic, $X_{n-k,n}$, where $X_{1,n} \leq \cdots \leq X_{n,n}$ are the order statistics of the sample; and (ii) the scaling function satisfies $a(t_I') \approx a(1/(1-F_n(\tau_I))) = a(n/k)$. Consequently, the peak values in the sample correspond to the top $k$ order statistics, $(X_{n-k+1,n},\ldots,X_{n,n})$, which can be used for inference on the parameter $\bftheta$. Accordingly, in what follows, the scale parameter $\sigma$ is taken to represent $a(n/k)$.
Second, we introduce an important property that a forecasting technique should satisfy when the focus is the prediction of tail events, called {\it tail equivalence}. This property ensures that if a value in the far tail of a distribution has a very small exceedance probability, the forecasting procedure assigns the same probability to it—up to an estimation error that is negligible compared to the probability itself. This prevents scenarios where the predictive density and the forecaster density appear close in some metric, despite exhibiting significantly different behavior in the tail.
\begin{defi}\label{def:tail_equiv}
Let $\tau_E\geq\tau_I$ such that $\tau^\star\to \tau_0\in[0,1]$ as $n\to\infty$, and let $\widehat{H}_{\tau_I}$ be an estimator of the predictive distribution $F_{t_I}^\star$. We say that the forecaster $\widehat{H}_{\tau_I}$ is $\tau^\star$-tail equivalent to the true predictive distribution $F_{t_I}^\star$, with associated quantile function $Q^\star_{\tau_I}$, if as $n\to\infty$
$$
\frac{1-\widehat{H}_{\tau_I}(Q^\star_{\tau_I}(1-\tau^\star))}{\tau^\star}=1+o_{\Prob}(1).
$$
As a consequence, the associated forecaster $\widehat{F}_n:=\tau_I + (1-\tau_I)\widehat{H}_{\tau_I}$ for tail events is $(1-\tau_E)$-tail equivalent to the true distribution $F$, meaning that, as  $n\to\infty$
$$
\frac{\widehat{F}_n(Q(\tau_E))}{1-\tau_E}=1+o_{\Prob}(1).
$$
\end{defi}
In simple terms, this definition ensures that $\widehat{H}_{\tau_I}$ ($\widehat{F}_n$) has the same tail of  $F_{t_I}^\star$ ($F$), which is a natural form of agreement to ask to any predictive procedure aimed at forecasting extreme events. In the following sections, we present both frequentist and Bayesian approaches for their estimation, establish the consistency of the proposed density estimators by quantifying their contraction rates, and discuss the implications of these results for statistical prediction.

\subsection{Frequentist approach}\label{sec:freq}
Here we describe a simple frequentist approach for the estimation of
the predictive distribution $F^\star_{t_E}$ and its density $f^*_{t_E}${,} corresponding to an extreme level $\tau_E\geq \tau_I$.
Let $T_{n,i}$, $i=1,2$ be suitable measurable functions. Let 
$$
\widehat\gamma_n=T_{n,1}(X_{n-k,n},..., X_{n,n}), \quad \widehat{\sigma}_n=T_{n,2}(X_{n-k,n},..., X_{n,n})
$$ 
be estimators of the tail index for $\gamma>-1/2$ and 
of the scaling parameter $\sigma>0$. Exploiting the GP-based approximations \eqref{eq:GP_predictive_dist} and \eqref{eq:GP_predictive_dens}, and plugging into them the estimators of their parameters we obtain the predictive distribution and density estimators
\begin{align*}
\widehat{H}_{\tau_E}^{{\text{(F)}}}(y)&=H_{\widehat{\bftheta}_n}
\left(
(y-X_{n-j,n}){\tau^\star}^{\widehat{\gamma}_n}-
{\widehat{\sigma}_n}\frac{1-{\tau^\star}^{\widehat{\gamma}_n}}{\widehat{\gamma}_n}
\right),\\ 
\widehat{h}_{\tau_E}^{{\text{(F)}}}(y)&=h_{\widehat{\bftheta}_n}
\left(
(y-X_{n-j,n}){\tau^\star}^{\widehat{\gamma}_n}-
{\widehat{\sigma}_n}\frac{1-{\tau^\star}^{\widehat{\gamma}_n}}{\widehat{\gamma}_n}
\right)
\tau^\star,
\end{align*}
where the superscript \lq\lq$\text{(F)}$" stands for \lq \lq frequentist" and $\widehat{\bftheta}_n=( \widehat{\gamma}_n,\widehat{\sigma}_n)^\top$.
Setting $\tau_E=\tau_I$ gives the estimators $\widehat{h}_{\tau_I}^{(F)}$ and $\widehat{H}_{\tau_I}^{(F)}$ of predictive distribution and density regarding peaks above an intermediate threshold. 
We now evaluate the accuracy of the proposed estimators. First, we introduce an absolute measure of accuracy for our density-based predictor, followed by an assessment of the reliability of the distribution-based predictor. Leveraging Corollary \ref{cor:newapprox}, we quantify the contraction rates of the density estimator $\widehat{h}_{\tau_E}^{{\text{(F)}}}$ toward the true densities $f_{t_E}^\star$ in terms of Hellinger distance. The following result formally establishes these rates.
\begin{prop}\label{pro:rate_Helli}
Assume that the conditions of Corollary \ref{cor:newapprox} are satisfied. Assume also that the following conditions are satisfied as $n\to \infty$:
\begin{inparaenum}\label{prop:cons}
\item\label{cond:est} 
$|\widehat{\gamma}_n-\gamma|=O_{\mathbb{P}}(1/\sqrt{k})$ and $ |\widehat{\sigma}_n/a(n/k)-1|=O_{\mathbb{P}}(1/\sqrt{k})$,
\item\label{cond:balance} $\sqrt{k}|A(n/k)|\to \lambda \in (0,\infty) $.
\end{inparaenum}
Then, for any sequence $\epsilon_n\downarrow0$ as $n\to\infty$, such that $k\epsilon_n^2 \to \infty$ and 
$\epsilon_n {w}_{\gamma}(1/\tau^\star) \to 0$ as $n\to\infty$, we have
		$$
		\hell(f_{t_E}^\star, 
		\widehat{h}_{\tau_E}^{{\text{(F)}}}
		)=O_{\Prob}\left(
		\sqrt{\epsilon_n {w}_{\gamma}(1/\tau^\star)}
		\right).
		$$
\end{prop}
Condition \ref{cond:est} of Proposition \ref{prop:cons} simply assumes that the estimator $\widehat{\bftheta}_n$ is consistent, with usual $1/\sqrt{k}$-rate. Condition \ref{cond:balance} is typically used to show that the estimator $\widehat{\bftheta}_n$, obtained on the basis of the POT approach, satisfies the asymptotic normality result
$$
\sqrt{k}\left(
\widehat{\gamma}_n-\gamma,
\frac{\widehat{\sigma}_n}{a(n/k)}-1
\right)\stackrel{d}{\to} \mathcal{N}(\bfbeta,\bfSigma),
$$ 
where $\bfbeta$ and $\bfSigma$ are certain bias vector and variance-covariance matrix \citep[see][Ch. 3--5]{dehaan+f06}. Here we additionally show that the 
estimator $\widehat{h}_{\tau_{E}}^{{\text{(F)}}}$ 
becomes close in Hellinger distance to the true predictive density $f^\star_{t_E}$, at a certain speed.
We provide two examples of classical estimators that satisfy the conditions of Proposition \ref{pro:rate_Helli} and comply with its results.
\begin{ex}\label{ex:MLE}
Under the assumptions of Proposition \ref{theo:hellrate} and condition \ref{cond:balance}  of  Proposition \ref{prop:cons}, with probability tending to 1 there exists a unique Maximum Likelihood (ML) estimator of $\bftheta$ given by
$$
\widehat{\bftheta}_n= \underset{ \bftheta \in\Theta}{\arg\max} \prod_{i=1}^{k}h_\bftheta \left(
X_{n-k+i,n}-X_{n-k,n}
\right)
$$
where 
$\Theta=(-1/2,\infty)\times(0,\infty)$, satisfying condition \ref{cond:est} of Proposition \ref{prop:cons}, see \citet[][Corollary 2.3]{dombry23}.
\end{ex}
\begin{ex}\label{ex:GPWM}
The Probability Weighted Moment (PWM) estimator of $\bftheta$ is defined as
$$
\widehat{\gamma}_n=1-\left(\frac{M_n^{(1)}}{2 M_n^{(2)}}-1\right)^{-1}, \quad \widehat{\sigma}_n=M_n^{(1)}\left(\frac{M_n^{(1)}}{2 M_n^{(2)}}-1\right)^{-1},
$$
where
$$
M_n^{(1)}=\frac{1}{k}\sum_{i=1}^{k}\left(X_{n-i+1,n}-X_{n-k,n}\right), \quad M_n^{(2)}=\frac{1}{k}\sum_{i=1}^{k}\frac{i}{k}\left(X_{n-i+1,n}-X_{n-k,n}\right).
$$
Under the assumptions of Proposition \ref{theo:hellrate} and condition \ref{cond:balance} of Proposition \ref{prop:cons}, and assuming further that $\gamma<1/2$, then PWM estimator satisfy condition \ref{cond:est} of Proposition \ref{prop:cons}, see Theorem 3.6.1 in \cite{dehaan+f06}.
\end{ex}
Exploiting the results of Proposition \ref{pro:rate_Helli}, next corollary establishes the accuracy of predictive regions for future 
peaks above the extreme threshold $t_E$.
\begin{cor}\label{cor:freqpred}
Assume that the conditions of Proposition \ref{pro:rate_Helli} are satisfied. Let $\mathcal{P}_{\tau_E}$ be a measurable set depending on sequence of excesses $X_{n-k+i,n}-X_{n-k,n}$, $i=1,\ldots,k$,    and satisfying for $\alpha \in (0,1)$,
\begin{equation}\label{eq:predregFreq}
	\int_{\mathcal{P}_{\tau_E}} \widehat{h}_{\tau_E}^{\text{(F)}}(x) \diff x =1-\alpha.
\end{equation}
For any sequence $\epsilon_n\downarrow0$ such that $\sqrt{k} \epsilon_n^2 \to \infty$ 
and $\epsilon_n {w_{\gamma}}(1/\tau^\star)\to0$, as $n\to\infty$, we have
		$$ 
		\Prob \left(
		X_{n+1}  \in \mathcal{P}_{\tau_E} | X_{n+1} > t_E
		\right)= 1-\alpha + O_{\Prob}\left(
		\sqrt{
			\epsilon_n {w_{\gamma}}(1/\tau^\star)
		}\right).
		$$
\end{cor}
A concrete example of how to select $k$, $\tau_E$ and $\epsilon_n$ so that the results in Corollary \ref{cor:freqpred} hold is reported next.
\begin{ex}\label{ex:fredpred}
Set $k=n^\delta \log^\eta(n)$, with $\delta \in (0,1)$, $\eta \in \Real$, and $\tau_E=1- \nu/n$, with $\nu=k^\zeta$ and $\zeta \in (0,1)$. Set $\epsilon_n=C_n/\sqrt{k}$, with $C_n$ being a sequence going to infinity arbitrarily slowly.  Then, Corollary \ref{cor:freqpred} 
guarantees that the prediction of peaks  over the true extreme quantile $Q(\tau_E)$ by means of the extreme regions $\mathcal{P}_{\tau_E}$, deduced from the predictive density $\widehat{h}_{\tau_E}^{\text{(F)}}$, implies a probability error that is not larger than $n^{-\delta \zeta/4}$, up to a logarithmic term proportional to $\sqrt{C_n} (\log n)^{-(\eta \zeta)/4}$.
\end{ex}
Note that Proposition \ref{prop:cons} implies $|\widehat{H}_{\tau_I}^{(\text{F})}(y)-F_{\tau_I}^\star(y)|=o_{\mathbb{P}}(1)$, uniformly in $y> t_I$, when $\tau_E = \tau_I$. We provide next some  
results on the relative accuracy of the estimators ${\widehat{H}}_{\tau_I}^{{\text{(F)}}}$ and $\widehat{F}_n^{\text{(F)}}: = \tau_I +(1-\tau_I) \widehat{H}_{\tau_I}^{
{\text{(F)}}}$ in estimating the probability of single events that are in the very far of the tail of the original distribution.
\begin{prop}\label{pro:consistency_freq_naive}
Under the assumptions of Corollary \ref{cor:newapprox}
and conditions \ref{cond:est}--\ref{cond:balance} of Proposition \ref{pro:rate_Helli}, for $\tau_E\geq \tau_I$ such that ${w_\gamma(1/\tau^\star)}=o(\sqrt{k})$ as $n\to \infty$, we have that 
the estimators ${\widehat{H}}_{\tau_I}^{\text{(F)}}$ and $\widehat{F}_n^{\text{(F)}}$ are $\tau^\star$ and $(1-\tau_E)$-tail equivalent to the true predictive distribution $F_{t_I}^\star$ and true underlying distribution $F$, as $n\to\infty$, respectively. Specifically,
 for any $\epsilon_n\downarrow0$ such that $k\epsilon_n^2\to\infty$  and $w_\gamma(1/\tau^\star)\epsilon_n\to 0$, as $n\to\infty$, 
 \begin{eqnarray*}
\frac{1-{\widehat{H}}_{\tau_I}^{\text{(F)}
	}(Q_{t_I}^\star(1-\tau^\star))}{\tau^\star}&=&1+O_\Prob(\epsilon_n w_\gamma(1/\tau^\star)),\\
\frac{1-{\widehat{F}_n^{\text{(F)}}(Q(\tau_E))}}{1-\tau_E}&=&1+O_\Prob(\epsilon_n w_\gamma(1/\tau^\star)).
 \end{eqnarray*}
\end{prop}

\subsection{Bayesian approach}\label{sec:bayes}
We present a Bayesian inferential procedure for the POT method, incorporating both classical prior formulations and empirical Bayes approaches. The method, summarized below, follows the framework detailed in \citet{dombry23}. This approach is based on a flexible prior specification for the GP distribution parameter $\bftheta$, with a density function of the form:
\begin{equation}\label{eq:prior_ist}
	\pi(\bftheta)= \pi_{\text{sh}}(\gamma) \pi_{\text{sc}}^{(n)}
	\left( \sigma\right), \quad \bftheta \in \Theta,
\end{equation}
where $ \pi_{\text{sh}}$ is a prior density on $\gamma$ and for each $n=1,2,\ldots$, $\pi_{\text{sc}}^{(n)}$ is a prior density on $\sigma$, whose expression may or may not depend on $n$, and where $\Theta$ is as in Example \ref{ex:MLE}.
To ensure that the resulting posterior distribution exhibits desirable theoretical guarantee, the following assumptions are made.
\begin{cond}\label{cond:prior}
The densities  $\pi_{\text{sh}}$ and $\pi_{\text{sc}}^{(n)}$ are such that:
\begin{inparaenum}
\item\label{cond:pisc} For each $n=1,2,\ldots$, $\pi_{\text{sc}}^{(n)}:\Real_+\to\Real_+$ and
\begin{inparaenum}
\item[(a.1)]\label{posit} there is a constant $\delta>0$ such that $\pi_{\text{sc}}^{(n)}(a_(n/k))a_(n/k)>\delta$ and for any constant $\eta>0$ there is $\epsilon>0$ such that 
$$
\sup_{ 1-\epsilon < \sigma <1+\epsilon} \left|
\frac{\pi_{\text{sc}}^{(n)}(a(n/k)\sigma)}{\pi_{\text{sc}}^{(n)}(a(n/k))}-1 \right|<\eta;
$$
\item[(a.2)]\label{piscbound} there is $C>0$ such that  ${\sup_{\sigma >0} \sigma a(n/k) \pi_{\text{sc}}^{(n)}(a(n/k)\sigma) \leq C}$;
\end{inparaenum}
Inequalities (a.1)-(a.2) hold with probability tending to $1$, for fixed $\delta, \eta, \epsilon, C$, if $\pi_{\text{sc}}^{(n)}$ is data-dependent.
\item\label{cond:pish}  $\pi_{\text{sh}}$ is a positive and continuous function at $\gamma_0$ such that: $\int_{-1/2}^0 \pi_{\text{sh}}(\gamma)\,\mathrm{d}\gamma<\infty$, $\sup_{\gamma>0}\pi_{\text{sh}}(\gamma)<\infty$.
\end{inparaenum}
\end{cond}
As discussed by \citet{dombry23}, there are two broad classes of priori distributions satisfying such conditions. One is the family of informative data dependent priors, where $\pi_{\text{sh}}(\gamma)=\pi(\gamma)$, with $\pi$ that is any continuous probability density on $(-1/2,\infty)$ bounded away from infinity, $\pi_{\text{sc}}^{(n)}(\cdot)=\pi(\cdot/\widehat{\sigma}_n)/\widehat{\sigma}_n$, where $\pi$ is an informative prior density on $(0,\infty)$ (Gamma, Inverse-Gamma, Weibull, Pareto, etc.) and $\widehat{\sigma}_n$ is consistent estimator of $a(n/k)$. Another one is the family of non-informative improper priors,  where $\pi_{\text{sh}}(\gamma)=\pi(\gamma)$, with $\pi$ that is any non-informative prior on $\gamma$ (e.g. uniform, maximal data information, and Jeffreys priors) and $\pi_{\text{sc}}^{(n)}(\sigma)=\pi(\sigma)\propto 1/\sigma$, i.e. a uniform distribution on $\log\sigma$.

Given a prior density $\pi$ as in \eqref{eq:prior_ist}, the posterior distribution of the GP distribution parameter $\bftheta$ is by Bayes rule 
\begin{equation}\label{eq:posterior_pot}
	\varPi_n(B)=\frac{\int_{B}\prod_{i=1}^k h_\bftheta(X_{n-i+1,n}-X_{n-k,n})\pi(\bftheta)\diff \bftheta}
	{\int_{\Theta}{\prod_{i=1}^k h_\bftheta(X_{n-i+1,n}-X_{n-k,n})}\pi(\bftheta)\diff \bftheta},
\end{equation}
for all measurable sets $B\subset \Theta$. Under the Bayesian paradigm, estimators of $F_{t_E}^{\star}$ and $f_{t_E}^{\star}$ are given for an extreme level $\tau_E\geq \tau_I$  by
\begin{align}
\nonumber \widehat{H}_{{\tau_E}}^{{\text{(B)}}}(y)&=\int_{\Theta} H_{\bftheta}
\left(
(y-X_{n-j,n}){\tau^\star}^{\gamma}-
\sigma\frac{1-{\tau^\star}^{\gamma}}{\gamma}
\right) \diff \varPi_n (\bftheta),\\ 
\label{eq:Bayes_predictive_den} \widehat{h}_{{\tau_E}}^{{\text{(B)}}}(y)&=\int_{\Theta}  h_{\bftheta}
\left(
(y-X_{n-j,n}){\tau^\star}^{\gamma}-
{\sigma}\frac{1-{\tau^\star}^{\gamma}}{\gamma}
\right)\tau^\star \diff \varPi_n (\bftheta),
\end{align}
where the superscript \lq \lq $\text{(B)}$" stands for \lq \lq Bayesian".
Setting $\tau_E=\tau_I$ also gives the Bayesian estimators $\widehat{H}_{\tau_I}^{(B)}$ and $\widehat{h}_{\tau_I}^{(B)}$ of predictive distributions and densities for peaks above an intermediate threshold.
According to \citet[][Theorem 2.8]{dombry23}, the posterior distribution $\varPi_n$ concentrates on values $\bftheta\in\Theta$ such that both $|\gamma -\gamma_0|$ and $|\sigma/a_0(n/k) 
-1|$ approach zero as $n$ increases. Furthermore, \citet[][Theorem 2.12]{dombry23} establish that the Bayesian estimator $\widehat{H}^{{\text{(B)}}}_{\tau_I}$ consistently estimates $F^\star_{t_I}$ in the Wasserstein distance. However, the accuracy of Bayesian forecasters remains only partially understood, particularly in the context of predicting future peaks above intermediate thresholds. To address this gap, we provide a more comprehensive analysis. Ensuring consistency at the density level for peaks over a broader range of thresholds allows for more refined results than those currently available. In this regard, Proposition \ref{prop:hellcons} shows that the posterior distribution $\varPi_n$ asymptotically concentrates on values $\bftheta\in\Theta$ such that the corresponding GP density ${h}_{\bftheta,\tau_E}^{\star}$ closely approximates the true predictive densities $f_{t_E}^\star$ in Hellinger distance.
\begin{prop}\label{prop:hellcons}
Assume that Condition \ref{cond:prior}, the conditions of Corollary \ref{cor:newapprox} and conditions \ref{cond:balance} of Proposition \ref{prop:cons} are satisfied. Then, for any sequence $\epsilon_n \downarrow0$ such that  $\sqrt{k} \epsilon_n^2 \to \infty$
and $\epsilon_n {w}_{\gamma}\left(1/\tau^\star\right) \to 0$ as $n \to \infty$, there is a $R>0$ such that
$$	
\varPi_n\left(
\left\lbrace
\bftheta \in \Theta: \, \hell (h_{\bftheta,{\tau_E}}^{\star},f_{t_E}^\star) > \sqrt{\epsilon_n {w}_{\gamma}\left(1/\tau^\star)\right)}
\right\rbrace
\right) = O_{\mathbb{P}}\left(
e^{-R  k \epsilon_n^2 }
\right).
$$
\end{prop}
Furthermore, the above proposition provides the foundation for obtaining additional practically useful results. In particular, Corollary \ref{cor:bayespred} guarantees the absolute accuracy of the estimator $\widehat{h}_{{\tau_E}}^{{\text{(B)}}}$, as well as the reliability of the predictive regions derived from it.
\begin{cor}\label{cor:bayespred}
Assume that conditions of Proposition \ref{prop:hellcons} are satisfied. Let $\mathcal{P}_{\tau_E}$ be a measurable set depending on sequence of excesses $X_{n-k+i,n}-X_{n-k,n}$, $i=1,\ldots,k$, 
{and} satisfying for $\alpha \in (0,1)$
\begin{equation}\label{eq:predregBayes}
	\int_{\mathcal{P}_{\tau_E}} \widehat{h}_{{\tau_E}}^{\text{(B)}}(y) \diff y =1-\alpha.
\end{equation}
For any sequence $\epsilon_n\downarrow 0$ such that $\sqrt{k} \epsilon_n \to \infty$, 
$\epsilon_n w_{\gamma}\left(1/\tau^\star\right) \to 0$ and $\log\left(k w_{\gamma}^2 \left(1/\tau^\star\right) \right)=o(k\epsilon_n^2)$ {as $n\to\infty$}, we have
$$
\hell(\widehat{h}_{{\tau_E}}^{{\text{(B)}}}, f_{t_E}^\star)=O_{\Prob}\left( \sqrt{\epsilon_n w_{\gamma}\left(1/\tau^\star\right)}\right)
$$
and
$$
\Prob(X_{n+1}\in\mathcal{P}_{\tau_E} | X_{n+1} > t_E)= 1-\alpha + O_{\Prob}\left(\sqrt{\epsilon_n w_{\gamma}(1/\tau^\star)}\right).
$$
\end{cor}
Similar to the frequentist case, with the same choice of $k$, $\tau_E$, and $\epsilon_n$ as in Example \ref{ex:fredpred}, Corollary \ref{cor:bayespred}(ii) ensures that predictions based on the extreme region $\mathcal{P}_{\tau_E}$ derived from the posterior predictive density $\widehat{h}^{{\text{(B)}}}_{\tau_E}$ yield a small probability error for such forecasting tasks. Finally, we present results on the relative accuracy of the estimators ${\widehat{H}}_{\tau_I}^{\text{(B)}}$ and ${\widehat{F}_{n}^{\text{(B)}}} := \tau_I + (1-\tau_I) \widehat{H}{\tau_I}^{(B)}$ in estimating the probability of single events occurring in the far tail of the original distribution.
\begin{prop}\label{pro:consistency_Bayesian}
Under the assumptions of Corollary \ref{cor:newapprox}, Condition \ref{cond:prior} and condition \ref{cond:balance} of Proposition \ref{pro:rate_Helli}, for $\tau_E\geq \tau_I$ such that $w_{\gamma}(1/\tau^\star)=o(\sqrt{k{/\log(1/\tau^\star)}} )$ as $n\to \infty$, we have that 
the estimators ${\widehat{H}}_{\tau_I}^{(B)}$ and $\widehat{F}_n^{(B)}$ are $\tau^\star$- and $(1-\tau_E)$-tail equivalent to the true distribution $F_{t_I}^\star$ and $F$, as $n\to\infty$. Specifically,
for any $\epsilon_n\downarrow0$ such that $k\epsilon_n^2\to\infty$, $-\log(\epsilon_n w_{\gamma}(1/\tau^\star))=o(k \epsilon_n^2)$ and $w_\gamma(1/\tau^\star)\epsilon_n\to 0$, as $n\to\infty$, we have
\begin{eqnarray*}
\frac{1-{\widehat{H}}_{\tau_I}^{\text{(B)}}(Q_{t_I}^\star(1-\tau^\star))}{\tau^\star}&=&1+O_\Prob(\epsilon_n w_\gamma(1/\tau^\star)),\\
\frac{1-\widehat{F}_n^{\text{(B)}}(Q(\tau_E))}{1-\tau_E}&=&1+O_\Prob(\epsilon_n w_\gamma(1/\tau^\star)).
 \end{eqnarray*}
\end{prop}
\section{Tail risk assessment}\label{sec:tail_risk_assessment}
In this section, we explain how deriving appropriate point forecasts from the estimators of the predictive distribution for tail events, discussed in the previous section, enables the consistent estimation of widely-used extreme risk measures such as VaR and ES. This is an important achievement, as it shows that our proposed predictive distributions offer more comprehensive risk assessments than standard approaches based solely on extreme risk measures. Not only can they recover traditional risk measures, but their reach also extends beyond the latter by enabling the computation of predictive regions.

Standard inferential methods for assessing the extreme VaR in \eqref{eq:extreme_VaR} have been extensively discussed in the literature \citep[e.g.,][Ch. 4]{dehaan+f06}. These approaches share a common root, which is based on the following extrapolation formula
\begin{equation}\label{eq:est_extreme_quantile}
\widehat{Q}_n(\tau_E)=X_{n-k,n} + \widehat{\sigma}_n\frac{{\tau^\star}^{-\widehat{\gamma}_n}-1}{\widehat{\gamma}_n},
\end{equation}
where $(\widehat{\gamma}_n, \widehat{\sigma}_n)$ can be obtained with the ML, the PWM or other estimation methods. For a Bayesian estimation counterpart see \citet{dombry23}.  Since the $(1-\tau^\star)$-quantile of the conditional distribution $F_{t_I}^\star$ is equal to the quantile 
$Q(\tau_E)$, an alternative way of obtaining estimators based on
our predictive distributions is through the formula
$$
\widehat{Q}_n^{\text{(M)}}(\tau_E)=
({\widehat{H}}_{\tau_I}^{{\text{(M)}}})^\leftarrow ({1-}\tau^\star),
$$
where ``M=F'' or ``M=B'' in the case of a frequentist or a Bayesian forecaster, respectively.  The following proposition demonstrates that both frequentist and Bayesian point forecasts provide consistent estimators of the extreme VaR.
\begin{prop}\label{pro:eVaR_point_forecast}
Assume that the assumptions of Corollary \ref{cor:newapprox} and conditions \ref{cond:est}--\ref{cond:balance} of Proposition \ref{pro:rate_Helli} hold. 
\begin{inparaenum}
\item For all $\epsilon_n\downarrow0$ such that $k\epsilon_n^2\to\infty$ 
and $-\log(\tau^\star)=o(\sqrt{k})$, as $n\to\infty$, we have
$$
\frac{\widehat{Q}_n^{\text{(F)}}(\tau_E)-Q(\tau_E)}{a(n/k){\tau^\star}^{-\gamma} w_\gamma(1/\tau_\star)}=O_{\Prob}(\epsilon_n).
$$
\item[(ii)]Additionally to the above conditions assume also that Condition \ref{cond:prior} holds and
$-\log \tau^\star = o(k \epsilon_n^2)$, $-\log \epsilon_n =o(k\epsilon_n^2)$ {as} $n\to\infty$, 
then
$$
\frac{\widehat{Q}_n^{\text{(B)}}(\tau_E)-Q(\tau_E)}{a(n/k){\tau^\star}^{-\gamma} w_\gamma(1/\tau_\star)}=O_{\Prob}(\epsilon_n).
$$
\end{inparaenum}
\end{prop}
We focus on the extreme ES in \eqref{eq:extreme_ES}, assuming that $\gamma<1$. To justify the consistent estimation of this measure using our predictive distributions, we first prove that $E(\tau_E)$ can be approximated with increasing accuracy by the expected value of ${H}_{t_E}^\star$, i.e.
$$
E^\star(\tau_E)= \int y\, h_{t_E}^\star(y) \diff y.
$$
\begin{prop}\label{pro:eES_approximation}
Assume that the assumptions of Corollary \ref{cor:newapprox} hold. Let $\gamma<1$ and $\tau_E\geq \tau_I$. If $-\log(\tau^\star)=o(\sqrt{k})$ as $n\to\infty$, then
$$
\frac{E(\tau_E)-E^\star(\tau_E)}{a(n/k){\tau^\star}^{-\gamma} w_\gamma(1/\tau_\star)}=O(A(n/k)).
$$
\end{prop}
The previous result suggests that the following summary
$$
\widehat{E}_n^{(\text{M})}(\tau_E)=\int y\, \widehat{h}_{{\tau_E}}^{\text{(M)}}(y) \diff y{}
$$
of the predictive density estimator $\widehat{h}_{\tau_E}^{\text{(M)}}$ also aims to estimate the extreme ES. Its accuracy is established in the following result.
\begin{prop}\label{pro:eES_point_forecast}
Assume that the assumptions of Corollary \ref{cor:newapprox} and conditions \ref{cond:est}--\ref{cond:balance} of Proposition \ref{pro:rate_Helli} hold. 
\begin{inparaenum}
\item[(i)] For all $\epsilon_n\downarrow0$ such that $k\epsilon_n^2\to\infty$ as $n\to\infty$, if
$-\log(\tau^\star)=o(\sqrt{k})$ {as} $n\to\infty$, then we have
$$
\frac{E(\tau_E)-\widehat{E}^{(F)}_n(\tau_E)}{a(n/k){\tau^\star}^{-\gamma} w_\gamma(1/\tau_\star)}=O_{\Prob}(\epsilon_n).
$$
\item[(ii)] Additionally to the above conditions, assume  
that Condition \ref{cond:prior}\ref{cond:pisc} holds, that $  \text{supp}(\pi_{\text{sh}}) \subset (-1/2,1)$ contains the true value of $\gamma$,  $\int_{-1/2}^0 \pi_{\text{sh}}(\gamma)\diff \gamma <\infty$, $\int_{0}^{1}(1-\gamma)^{-1}\pi_{\text{sh}}(\gamma)\diff \gamma <\infty$
and $-\log \tau^\star = o(k \epsilon_n^2)$, $-\log \epsilon_n =o(k\epsilon_n^2)$, as $n\to\infty$. Then 
$$
\frac{E(\tau_E)-\widehat{E}^{(B)}_n(\tau_E)}{a(n/k){\tau^\star}^{-\gamma} w_\gamma(1/\tau_\star)}=O_{\Prob}(\epsilon_n).
$$
\end{inparaenum}
\end{prop}

\section{Statistical prediction in time series}\label{sec:time_series}
In this section we describe a simple way to predict tail events in a time series context. 
In order to obtain an accessible approach for forecasting extreme values conditionally on  tail events that incorporates information on the past of the time series in a dynamic way, we focus on the special subclass of time series
$
Y_i = \mu_i + \xi_i\varepsilon_i,
$
where $\mu_i$ and $\xi_i$ are random terms which are dependent on a subset of past observations $\boldsymbol{S}_i \subseteq (Y_{i-1}, Y_{i-2}, \ldots)$, and $(\varepsilon_i)$ is a strictly stationary sequence of iid innovations, whose distribution $F$ satisfies the domain of attraction condition in Section \ref{sec:background}.  A simple example is the AR$(p)$, where $\boldsymbol{S}_i=(Y_{i-1}, \ldots, Y_{i-p})$, $\mu_i= \sum_{j=1}^p \phi_i Y_{i-j}$ and $\xi_i\equiv 1$, for suitable coefficients $\phi_1, \ldots, \phi_p$. Here we focus on a wider that includes popular heteroskedastic location-scale time series models such as ARMA and ARMA/GARCH.

Given a sample $\boldsymbol{Y}n := (Y_1, \ldots, Y_n)$, if the goal is to analyze the extremal behavior of an ``out-of-sample'' random variable $Y{n+1}$, and assuming we knew the distribution of the future innovation $\varepsilon_{n+1}$, along with the specific relationship linking $\mu_{n+1}$ and $\xi_{n+1}$ to past values $\boldsymbol{S}n$ (which could represent the entire past sequence, as in a causal ARMA process), then prediction could be made using the exact form $Y_{n+1} = \mu_n + \xi_n \varepsilon_{n+1}$. Since this is not the case in practice and since innovations are not observable, we proceed as follows.
For $i=1, \ldots,n$, we  first obtain estimators $\widehat{\mu}_{i}^{(n)}$ and $\widehat{\xi}_i^{(n)}$ of $\mu_i$ and $\xi_i$, then we define residuals 
$
\widehat{\varepsilon}_{i}^{(n)}=(Y_i-\widehat{\mu}_{i}^{(n)})/\widehat{\xi}_i^{(n)}.
$
We regard the estimators  $(\widehat{\varepsilon}_{i}^{(n)})$ of the unobserved $(\varepsilon_i)$ as 
a sequence of (approximately) independent variables and we focus on prediction of the ``out-of-sample'' estimator $\widehat{\varepsilon}_{n+1}^{(n)}$ of the error $\varepsilon_{n+1}$. 
Precisely, we study the conditional distribution $\Prob(\widehat{\varepsilon}_{n+1}^{(n)} \leq z \mid \widehat{\varepsilon}_{n+1}^{(n)}  >\widehat{\varepsilon}_{n-k,n}^{(n)}, \widehat{\bfvarepsilon}_n)$, for a large value $z> \widehat{\varepsilon}_{n-k,n}^{(n)}$, where $\widehat{\bfvarepsilon}_n=(\widehat{\varepsilon}_{1}^{(n)},\ldots, \widehat{\varepsilon}_{n}^{(n)})$ and $\widehat{\varepsilon}_{n-k,n}^{(n)}$ is the $(n-k)$th order statistic, which is representative of the predictive distribution of future extreme innovations  $\Prob(\varepsilon_{n+1} \leq z \mid \varepsilon_{n+1} > Q(\tau_E), \bfvarepsilon_n)$, where $\bfvarepsilon_n=(\varepsilon_1,\ldots,\varepsilon_n)$ and $ Q(\tau_E)$ is the extreme quantile of the innovations distribution. Afterwards, we apply the methodology described in Section \ref{sec:strong_results} to estimate the corresponding predictive density, for different extreme levels $\tau_E\geq \tau_I$. Let $\widehat{h}_{{\tau_E}}^{{\text{(M)}}}(z)$ be the estimator of the predictive distribution of $\varepsilon_{n+1}$, where again ``M=F'' stands for frequentist  and ``M=B'' stands for Bayesian, denoting the method used for the estimation of the density. 
Given a value $z$ at the innovation scale, and the transformation $y = \widehat{\mu}_{n+1}^{(n)}+ \widehat{\xi}_{n+1}^{(n)}z$, then an estimator of the predictive density of $Y_{n+1}$, given that it exceeds the extreme  quantile $Q_{n+1}(\tau_E):=\mu_{n+1}+\xi_{n+1}Q(\tau_E)$ of its conditional distribution on the past of the time series, is given by
%
$$
\widehat{f}_{\tau_E}^{\text{(M)}}(y\mid \boldsymbol{Y}_{n})= \widehat{h}_{{\tau_E}}^{{\text{(M)}}}((y-\widehat{\mu}_n^{(n)})/ \widehat{\xi}_{n+1}^{(n)})/ \widehat{\xi}_{n+1}^{(n)},
$$
where $Q(\tau_E)$ is the extreme quantile of the innovations distribution. The corresponding distribution provides an estimator of the predictive distribution $\Prob(Y_{n+1}\leq y \mid Y_{n+1}>Q_{n+1}(\tau_E), \boldsymbol{S}_n)$, whose quantile and density functions are hereafter denoted for simplicity by $Q_{t_{n+1}}^\star(\tau)$ and $f_{t_{n+1}}^\star(y)$, where $t_{n+1}= Q_{n+1}(\tau_E)$ stands for the past data-dependent threshold to be exceeded. Finally, for $\tau_E=\tau_I$, let 
$$
\widehat{F}_{n+1}^{(M)}(y):= \tau_I+ (1-\tau_I) \widehat{F}_{\tau_I}^{(M)}(y \mid \bfY_n),
$$
with $y>\widehat{Q}_{n+1}(\tau_I )$, be an estimator of the true conditional  predictive distribution $\mathbb{P}(Y_{n+1}\leq y \mid \boldsymbol{S}_{n})$ for a large value $y>Q_{n+1}(\tau_I )$ given past observations, where $\widehat{Q}_{n+1}(\tau)= \widehat{\mu}_{n+1}^{(n)}+\widehat{\xi}_{n+1}^{(n)} \widehat{Q}_n^{\text{(M)}}(\tau)$ is an estimator of $Q_{n+1}(\tau )$ and 
$\widehat{Q}_n^{(\text{M})}(\tau)$ is an estimator of innovations quantile $Q(\tau)$ as in Section \ref{sec:tail_risk_assessment}, with $\tau \in [\tau_I,1)$.
Next are the minimal conditions required {to establish asymptotic accuracy for the introduced estimators.
\begin{cond}\label{cond:onestepahead}
There are positive sequences $m_n =o(1)$ and $l_n=o(1)$ such that
\begin{inparaenum}
\item \label{cond:bigOh}	
$\left|\frac{\mu_{n+1}- \widehat{\mu}_{n+1}^{(n)}}{\widehat{\xi}_{n+1}^{(n)}}
\right|=O_{\mathbb{P}}(m_n)$ and $\left|
\frac{\xi_{n+1}}{\widehat{\xi}_{n+1}^{(n)}}-1
\right|=O_{\mathbb{P}}(l_n);
$
\item \label{cond:seq}
$\delta_n:=l_n \frac{|t_I + a(n/k) ((1/\tau^\star)^\gamma-1)/\gamma| }{a(n/k)(1/\tau^\star)^\gamma} +\frac{m_n}{a(n/k) (1/\tau^\star)^\gamma}=o(1).
$
\end{inparaenum}
\end{cond}
\begin{prop}\label{pro:rate_Helli_TS}
We work under conditions of Corollary \ref{cor:newapprox}, \ref{cond:balance} of Proposition \ref{pro:rate_Helli} and Condition \ref{cond:onestepahead}. Assume that, for any $C_n \to \infty$ with $C_n =o(\sqrt{k})$,
\begin{equation}\label{eq:rateest}
|\widehat{\gamma}_n-\gamma|=O_{\mathbb{P}}(C_n/\sqrt{k}), \quad  |\widehat{\sigma}_n/a(n/k)-1|=O_{\mathbb{P}}(C_n/\sqrt{k}),
\end{equation}
and $\sqrt{k}(\widehat{\varepsilon}^{(n)}_{n-k,n}-t_I)=O_\mathcal{P}(n/k)$.
Then, for any $\epsilon_n \downarrow0$ such that $\sqrt{k}\epsilon_n \to \infty$ and $\epsilon_n {w}_{\gamma}(1/\tau^\star) \to 0$, as $n\to\infty$:
\begin{inparaenum}
\item[(i)]
$
\hell \left(f_{t_{n+1}}^\star, 
\widehat{f}_{\tau_E}^{\text{(F)}}(\, \cdot \, \mid  \bfY_{n})
\right)=O_{\Prob}(
\sqrt{\epsilon_n {w}_{\gamma}(1/\tau^\star) + \delta_n } + l_n),
$ with $t_{n+1}=Q_{n+1}(\tau_E)$;
\item[(ii)] For any region  $\mathcal{P}_{\tau_E}$ satisfying \eqref{eq:predregFreq}, with $\mathcal{R}_{\tau_E}= \widehat{\mu}_{n+1}^{(n)} + \widehat{\xi}_{n+1}^{(n)} \mathcal{P}_{\tau_E}$, we have
$$
\Prob\left(Y_{n+1}  \in \mathcal{R}_{\tau_E} \mid Y_{n+1} > Q_{n+1}(\tau_E), \boldsymbol{S}_n \right)=  1-\alpha + O_{\Prob}(
\sqrt{\epsilon_n {w}_{\gamma}(1/\tau^\star) +\delta_n} + l_n);
$$
\item[(iii)] If $\sqrt{k} m_n =O(a(n/k)(1/\tau^\star)^\gamma w_\gamma(1/\tau^\star))$ and  $\sqrt{k}l_n t_E =O(a(n/k)(1/\tau^\star)^\gamma w_\gamma(1/\tau^\star))$, then, with $t_{n+1}=Q_{n+1}(\tau_I)$, we have tail equivalences
			\begin{eqnarray*}
				\frac{1-{\widehat{F}}_{\tau_I}^{\text{(F)}}(Q_{t_{n+1}}^\star(1-\tau^\star)\mid \bfY_{n})}{\tau^\star}&=&1+O_\Prob(\epsilon_n w_\gamma(1/\tau^\star)),\\
				\frac{1-{\widehat{F}_{n+1}^{\text{(F)}}(Q_{n+1}(\tau_E)  )}}{1-\tau_E}&=&1+O_\Prob(\epsilon_n w_\gamma(1/\tau^\star)).
			\end{eqnarray*}
		\end{inparaenum}
	\end{prop}
In the special case that $\widehat{\bftheta}_n$ is the maximum likelihood estimator, a sufficient condition for \eqref{eq:rateest} to be valid is that  as $n \to \infty$
\begin{equation}\label{eq:residualrate}
\max_{0\leq i \leq k}  \frac{|
		\widehat{\varepsilon}_{n-i+k,n}^{(n)}
		-\varepsilon_{n-i+k,n}|}{a(n/k)} =o_{\mathbb{P}}(1/\sqrt{k}).
	\end{equation}
This condition is useful for GP likelihood-based inference in general, as shown by next results regarding the Bayesian method.
\begin{prop}\label{prop:hellconsTS}
Let $\varPi_n$ be as in \eqref{eq:posterior_pot}, with $\widehat{\varepsilon}_{n-i+k,n}^{(n)}$ in place of $X_{n-k+i,n}$. Work under Condition \eqref{eq:residualrate}, \ref{cond:prior}, conditions of Corollary \ref{cor:newapprox} and condition \ref{cond:balance} of Proposition \ref{pro:rate_Helli}. Then, the results in Proposition \ref{prop:hellcons} are valid in this context.
\end{prop}
\begin{cor}\label{cor:bayespred_TS}
Work under Condition \ref{cond:onestepahead} and conditions of Proposition \ref{prop:hellconsTS}.
Then, for any $\epsilon_n \downarrow0$ such that $\sqrt{k} \epsilon_n \to \infty$, $\epsilon_n w_{\gamma}\left(1/\tau^\star\right) \to 0$ and $\log\left(k w_{\gamma}^2 \left(1/\tau^\star\right) \right)=o(k\epsilon_n^2)$ as $n\to\infty$, we have:
\begin{inparaenum}
\item 
$
\hell\left(
f_{t_{n+1}}^\star,
\widehat{f}_{\tau_E}^{(B)}(\, \cdot \mid \boldsymbol{Y}_{n})
			\right)=O_{\Prob}(\sqrt{\epsilon_n w_{\gamma}\left(1/\tau^\star\right)
		+ \delta_n }+  l_n ),
			$ with $t_{n+1}=Q_{n+1}(\tau_E)$;
		\item For any region  $\mathcal{P}_{\tau_E}$  that satisfies \eqref{eq:predregBayes}, with $\mathcal{R}_{\tau_E}= \widehat{\mu}_{n+1}^{(n)} + \widehat{\xi}_{n+1}^{(n)} \mathcal{P}_{\tau_E}$,  we have
			$$
			\Prob\left(Y_{n+1}  \in \mathcal{R}_{\tau_E} \mid Y_{n+1} > Q_{n+1}(\tau_E), \boldsymbol{S}_n\right)=  1-\alpha + O_{\Prob}(\sqrt{\epsilon_n w_{\gamma}\left(1/\tau^\star\right)
			+\delta_n	}+  l_n );
			$$
\item If  $\sqrt{k} m_n =O(a(n/k)(1/\tau^\star)^\gamma w_\gamma(1/\tau^\star))$ and $\sqrt{k}l_n t_E =O(a(n/k)(1/\tau^\star)^\gamma w_\gamma(1/\tau^\star))$, then, with $t_{n+1}=Q_{n+1}(\tau_I)$,
we have tail equivalences
			\begin{eqnarray*}
				\frac{1-{\widehat{F}}_{\tau_I}^{\text{(B)}}(Q_{t_{n+1}}^\star(1-\tau^\star)\mid \boldsymbol{Y}_{n})}{\tau^\star}&=&1+O_\Prob(\epsilon_n w_\gamma(1/\tau^\star)),\\
				\frac{1-{\widehat{F}_{n+1}^{(B)}(Q_{n+1}(\tau_E ))}}{1-\tau_E}&=&1+O_\Prob(\epsilon_n w_\gamma(1/\tau^\star)).
			\end{eqnarray*}
		\end{inparaenum}
	\end{cor}
We provide some examples of time series models and estimators $\widehat{\mu}_{n+1}$ and $\widehat{\xi}_{n+1}$ complying with Condition \ref{cond:onestepahead} and \eqref{eq:residualrate}.
	\begin{ex} In the case of ARMA$(p,q)$ model
	$Y_i = \sum_{j=1}^p \phi_j Y_{i-j}+ \sum_{j=1}^q \psi_j \varepsilon_{i-j} + \varepsilon_i$ and heavy tailed innovations, with $\gamma <1/2$, $\sqrt{n}$-consistent estimators $(\widehat{\phi}_j)_{j=1}^p, (\widehat{\psi}_j)_{j=1}^q$ are available (e.g., least square), yielding fitted values  for $i=1,2,\ldots,n$ 
	$$
	\widehat{\mu}_i^{(n)} = \sum_{j=1}^p \widehat{\phi}_{j} Y_{i-j}+ \sum_{j=1}^q \widehat{\psi}_j \widehat{\varepsilon}_{i-j}^{(n)}, \quad \widehat{\sigma}_i^{(n)} \equiv 1, 
	$$
	and residuals $\widehat{\varepsilon}_{\max(p,q)-q+1}^{(n)}= \ldots =\widehat{\varepsilon}_{\max(p,q)}^{(n)}=0$, $\widehat{\varepsilon}_{i}^{(n)}=Y_i-\sum_{j=1}^p \widehat{\phi}_{j} Y_{i-j}- \sum_{j=1}^q \widehat{\psi}_j\widehat{\varepsilon}_{i-j}^{(n)}$, $i=\max(p,q)+1, \ldots,n$. Under regularity conditions in \citet[][section 3.4.1]{g+s+u2021}, $\max_{1 \leq i \leq n} |\varepsilon_i- \widehat{\varepsilon}_i^{(n)}|=O_{\mathbb{P}}(1)$. Thus, if innovations distribution meets Conditon \ref{cond:SO},  a sufficient condition for \eqref{eq:residualrate} is $k^{1/2+\gamma}=o(n^\gamma)$.  Moreover, Condition \ref{cond:onestepahead}\ref{cond:bigOh} holds with $l_n\equiv 0$ and $m_n=1/\sqrt{n}$. Hence Condition \ref{cond:onestepahead}\ref{cond:seq} and the additional assumptions at point (iii) of Proposition \ref{prop:hellconsTS} and Corollary \ref{cor:bayespred_TS} are satisfied for any $\nu \leq k=o(n)$.
	\end{ex}
	\begin{rem}
	 The proposed methodology immediately extend to the case where $\boldsymbol{S}_{t}$ includes not just a subset of past observations $Y_{t-1}, Y_{t-2}, \ldots,$ but also exhogenous explanatory variables as, e.g., in the ARMAX model framework. Under Condition \ref{cond:onestepahead} and equations \eqref{eq:rateest}--\eqref{eq:residualrate}, our theoretical results are also valid in such a case.
	\end{rem}
}

\section{Real data analysis}\label{sec:real_analysiss}
\subsection{Prediction of extreme temperatures in Milan}\label{sec:real_temperatures}

We analyze the summer temperatures in Milan, Italy, from 1991 to 2023, focusing on the Daily Maximum Temperatures (DMT) recorded between June and September. For simplicity in forecasting, we assume that DMT are independent, even though they exhibit temporal dependence in practice. Given that we are only examining summer temperatures, there is no seasonal effect in the data. In 2003, a massive heatwave struck Europe, breaking temperature records across the continent. In Milan, on August 11th, the highest temperature ever recorded reached 38.3°C. The heat was exacerbated by high humidity, a characteristic of large urban cities like Milan. This analysis aims to predict temperature peaks above certain thresholds and assess the likelihood of reaching extreme temperature levels, similar to or exceeding the historical record.
%

After removing the missing values, we are left with $n=3140$ DMT. 
We use an effective sample size of $k=169$ and estimate the corresponding intermediate threshold $t_I$ by the order statistic $x_{n-k,n}=34$°C so that there are approximately $k/n\cdot 100\approx 5.4\%$ of higher temperatures in the sample. Such temperatures are then used to fit the GP distribution using the MLE, GPWM \citep[e.g.][Ch. 3]{dehaan+f06} and the Bayesian approach \citep{dombry23}, where a data-dependent prior is used for $\sigma$ and a data-independent prior is used for $\gamma$. In the sequel we abbreviate Asymmetric 95\% Credible Intervals and Predictive Intervals by A95\%CI and A95\%PI, respectively. 

\begin{figure}[t!]
	\centering
	\includegraphics[width=0.27\textwidth, page=2]{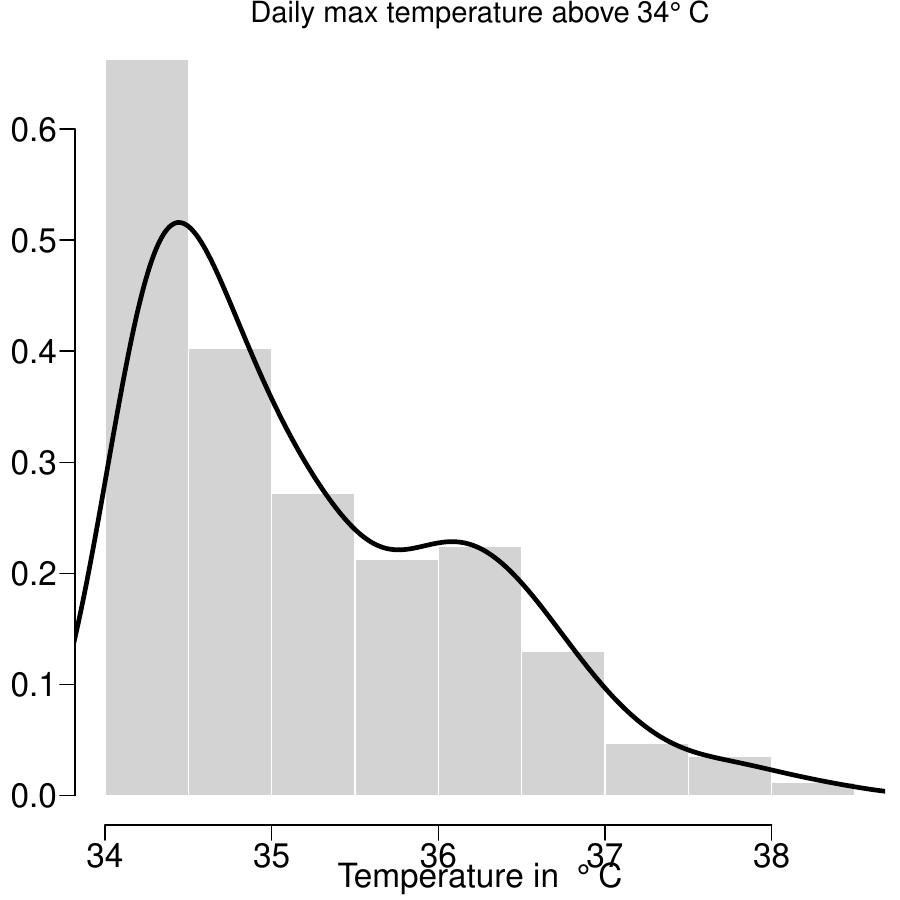}
	\includegraphics[width=0.27\textwidth, page=3]{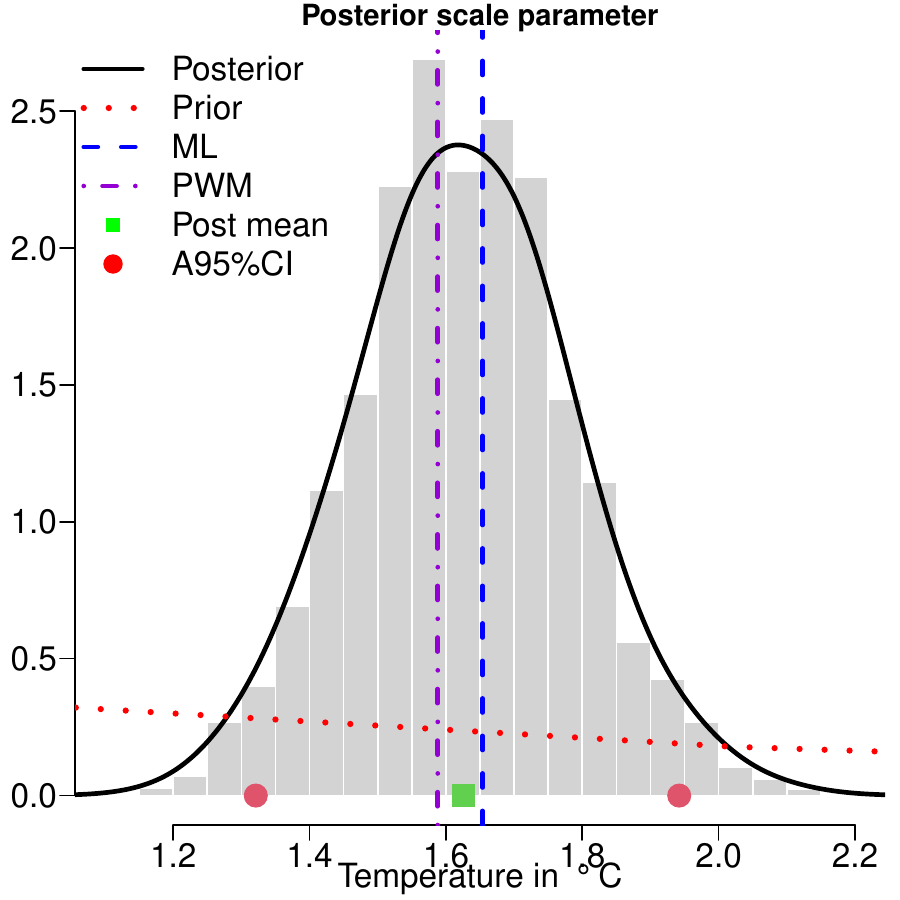}
	\includegraphics[width=0.27\textwidth, page=4]{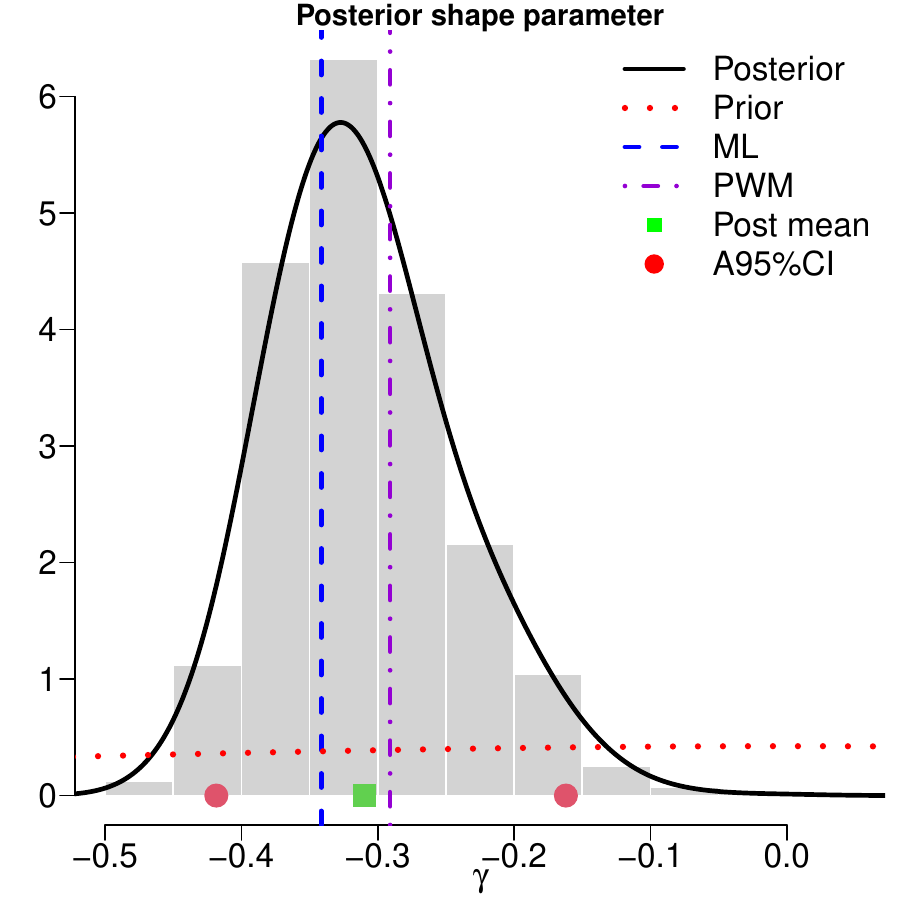}\\
	\includegraphics[width=0.27\textwidth, page=5]{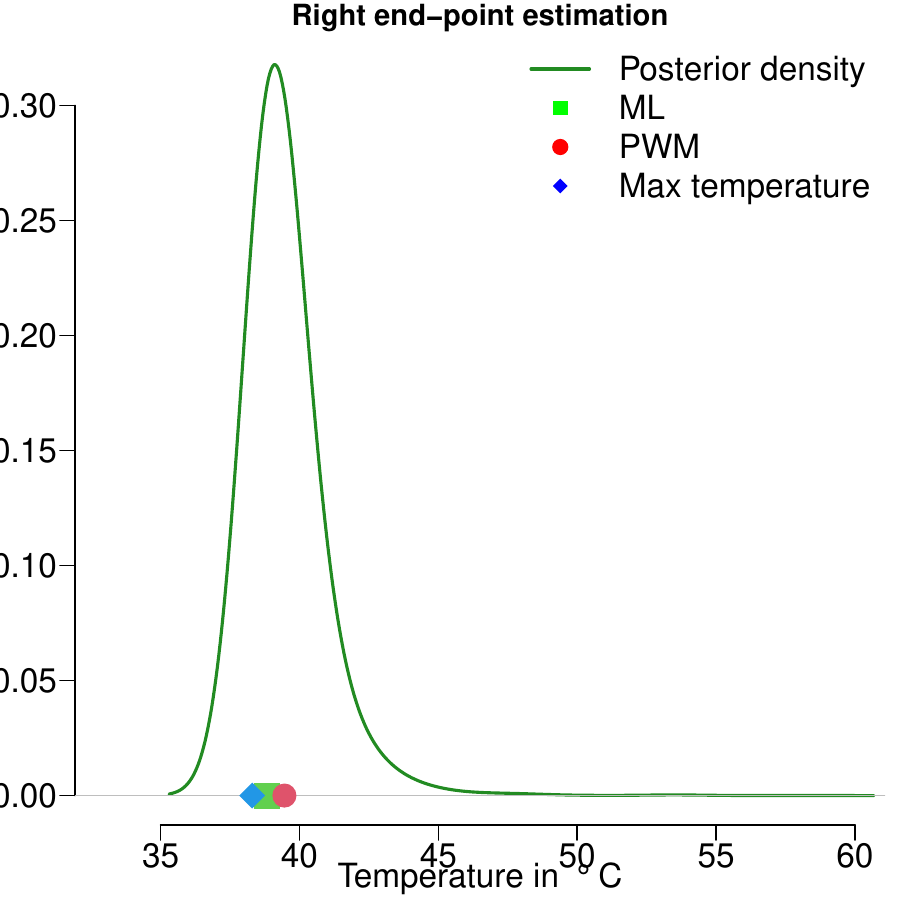}
	\includegraphics[width=0.27\textwidth, page=6]{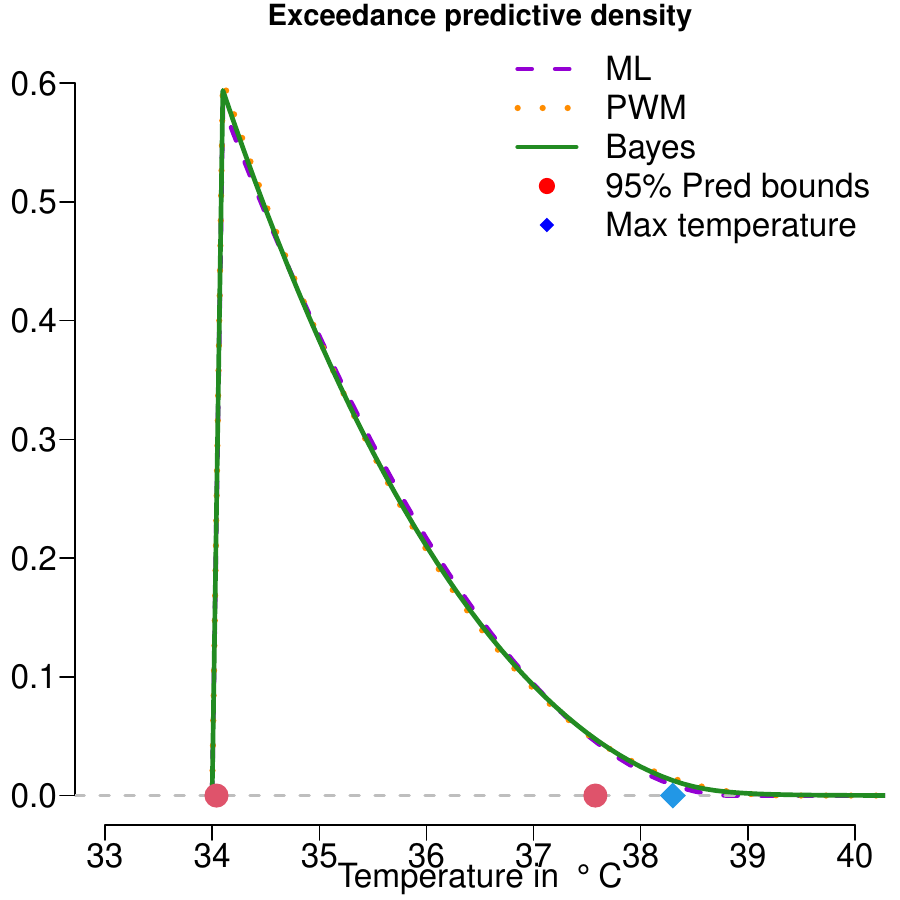}
	\includegraphics[width=0.27\textwidth, page=7]{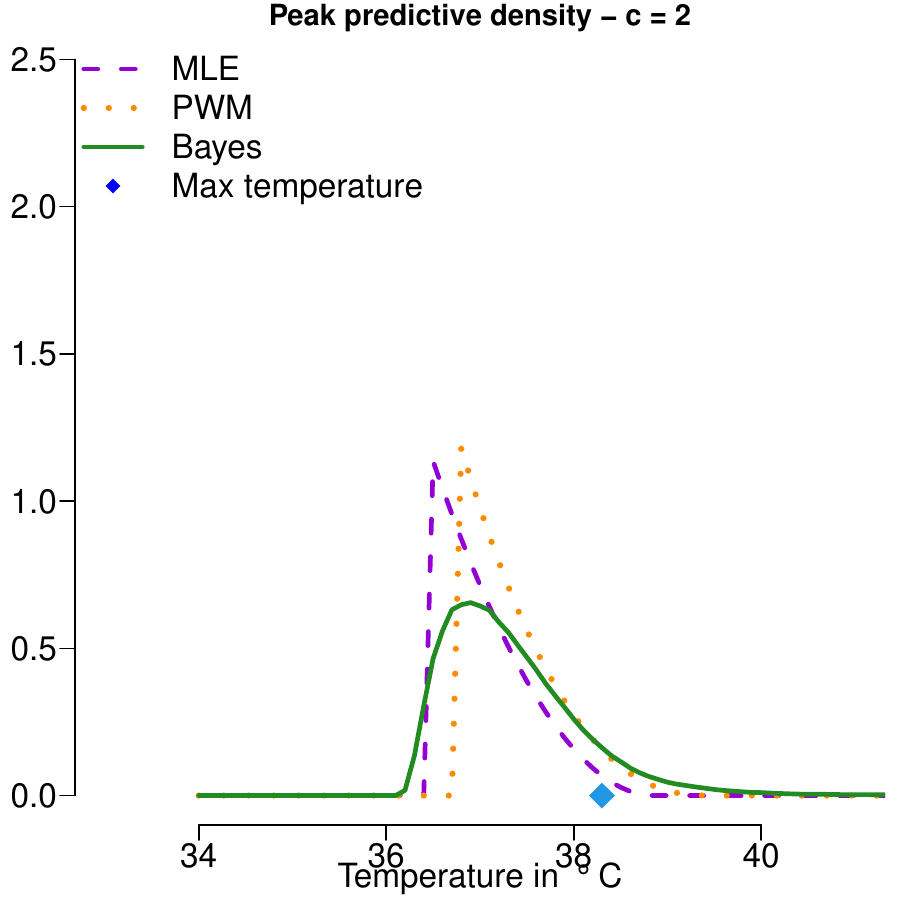}\\
	\includegraphics[width=0.27\textwidth, page=8]{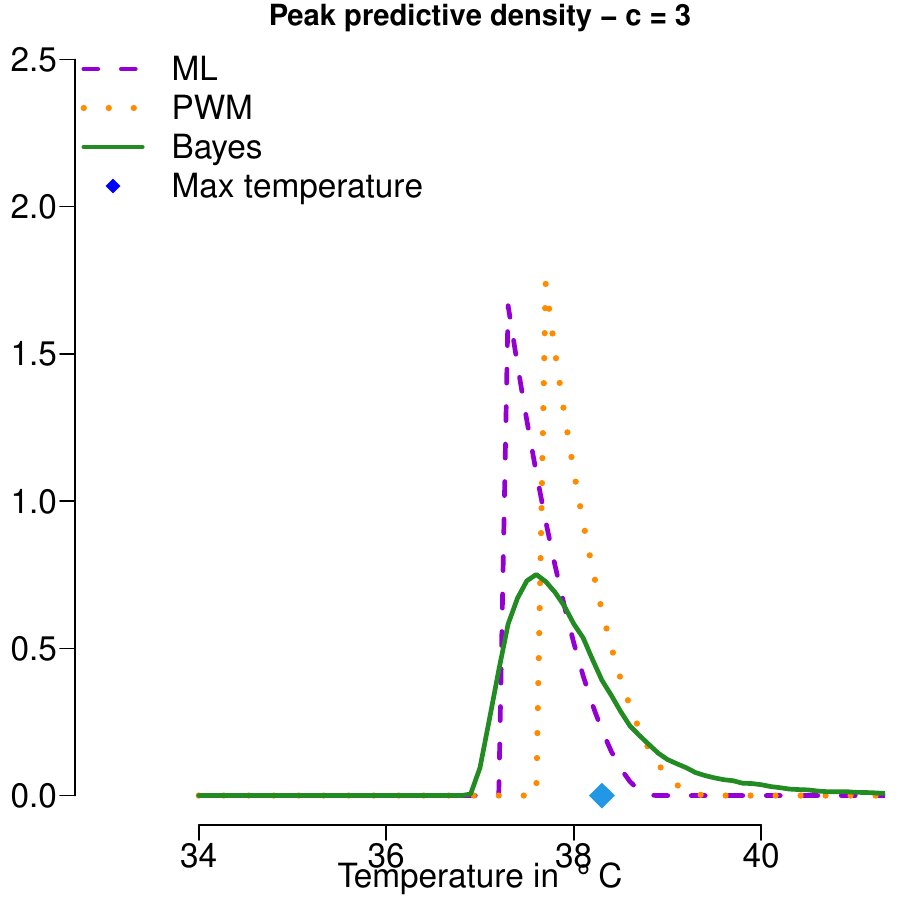}
	\includegraphics[width=0.27\textwidth, page=9]{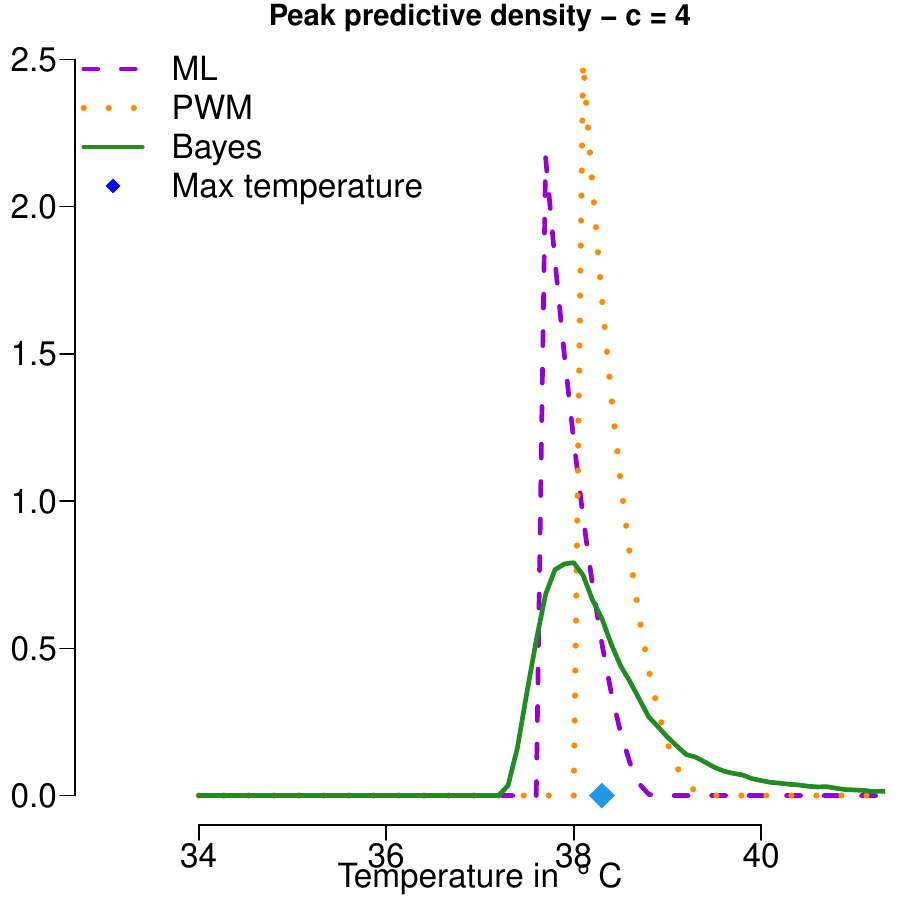}
	\includegraphics[width=0.27\textwidth, page=10]{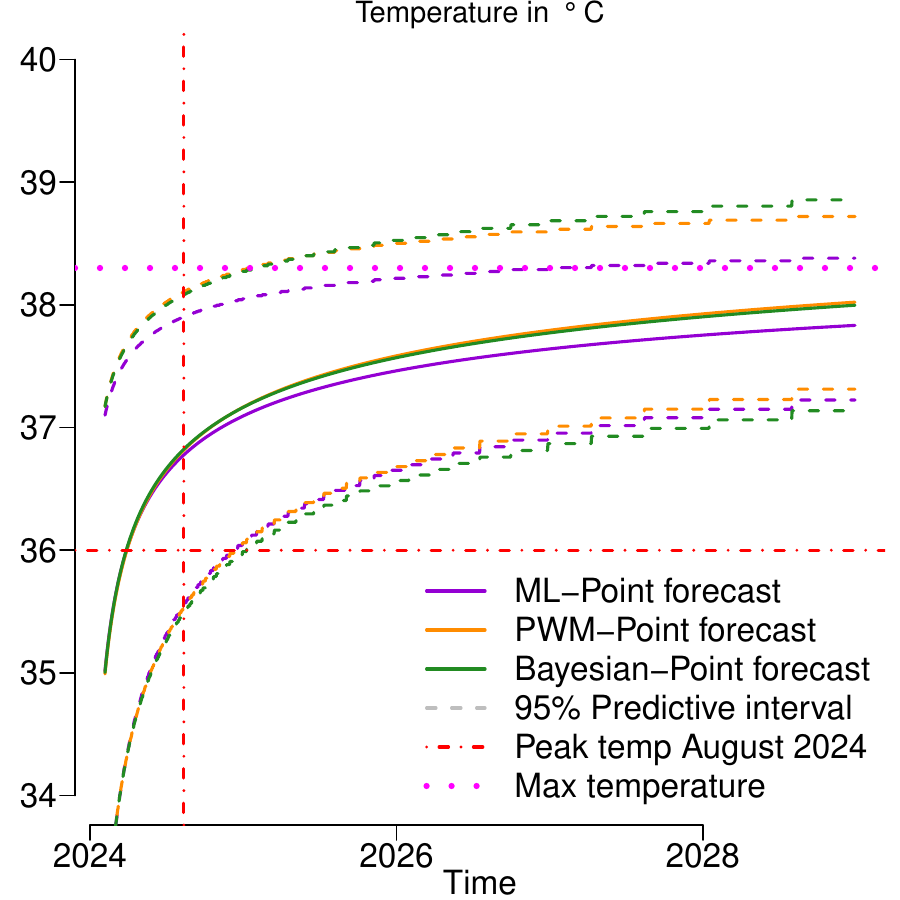}
	\caption{The top-left panel shows the empirical distribution of peak temperatures. The top-center and top-right panels display the posterior (black) and prior (red) densities of $\sigma$ and $\gamma$, with green squares and red circles indicating the posterior mean and A95\%CI. Blue dashed and violet dot-dashed lines mark the ML and GPWM estimates. The middle-left panel reports the posterior density of the right endpoint, alongside ML and GPWM estimates. From middle- to bottom-center, predictive densities of threshold exceedances for $c = 1$ to $4$ are shown for ML (violet), GPWM (orange), and Bayesian (green) methods. The maximum recorded temperature is marked with a blue diamond. The bottom-right panel shows five-year return level forecasts beyond 2023 using ML, PWM, and Bayesian methods, with dotted lines for A95\%PI. Red dot-dashed and magenta dotted lines mark the August 2024 and historical peak temperatures, respectively.}\label{fig:real_analysis}
\end{figure}

The ML and GPWM estimates of $(\gamma, \sigma)$ are (-0.34,1.65) and (-0.29,1.59), respectively. The top-left and top-middle panels of Figure \ref{fig:real_analysis} show the empirical posterior densities of $(\gamma, \sigma)$, derived from a sample of $M=20$th values drawn from the posterior \citep[see][for details on the MCMC algorithm used]{padoan24b}. Posterior means are (-0.31,1.63), with A95\%CI of [-0.42,-0.16] for $\gamma$ and [1.32,1.94] for $\sigma$, calculated using the posterior quantiles. These results suggest that the distribution $F$ of summer DMT is short-tailed with a finite upper endpoint $x_E$, as expected. To explore the potential for higher future temperatures, understanding how far they could plausibly rise in comparison with those observed in the sample, we estimate $x_E$. The ML and GPWM estimates are 38.84°C and 39.46°C, respectively. The top-right panel of Figure \ref{fig:real_analysis} shows the empirical posterior density of $x_E$, with mean 39.46°C and A95\%CI [38.43°C, 42.54°C].
\begin{table}[t!]
\centering
\caption{A95\%PI computed using the threshold $X_{n-k,n}$ (4th row), are reported for the ML (left section), GPWM (middle section), and Bayesian (right section) methods. A95\%PI based on ${t_E}=Q(\tau_E)$ are provided (3rd, 6th, 9th columns, and rows 6–8), where $\tau_E$ and $Q(\tau_E)$ are approximated using ML (1st, 2nd columns) and GPWM (4th, 5th columns) estimates, and posterior samples whose A95\%CI are derived (7th, 8th columns).}
	{\scriptsize
		\begin{tabular}{ccc|ccc|ccc}
			\toprule
			\multicolumn{9}{c}{Method}\\
			\midrule
			\multicolumn{3}{c}{ML} & \multicolumn{3}{c}{GPWM} &\multicolumn{3}{c}{Bayes}\\
			\midrule
			$\tau_I$ & $X_{n-k,n}$ & A95\%PI & $k/n$ & $X_{n-k,n}$ & A95\%PI & $\tau_I$ & $X_{n-k,n}$ & A95\%PI\\
			\midrule
			94.6  & 34.0 & [34.1, 37.5] & 94.6 & 34.0 & [34.1, 37.6] & 94.6 & 34.0 & [34.1, 37.6]\\
			\midrule
			$\widehat{\tau}_E$ & $\widehat{Q}(\widehat{\tau}_E)$ & A95\%PI & $\widehat{\tau}_E$ & $\widehat{Q}(\widehat{\tau}_E)$ & A95\%PI & $\tau_E$-A95\%CI & $Q(\tau_E)$-A95\%CI & A95\%PI\\
			\midrule
			99.293 & 36.4 & [36.4, 37.9] & 99.503 & 36.7 & [36.7, 38.3] & [98.973, 99.925] & [36.2, 38.2] & [36.4, 39.1]\\ 
			99.784 & 37.2 & [37.2, 38.2] & 99.877 & 37.6 & [37.6, 38.7] & [99.610, 99.994] & [36.9, 39.8] & [37.0, 40.2]\\
			99.907 & 37.6 & [37.6, 38.4] & 99.954 & 38.1 & [38.3, 38.8] & [99.804, 99.999] & [37.2, 40.5] & [37.4, 41.6]\\
			\bottomrule
		\end{tabular}
	}
	\label{tab:_real_analysis}
\end{table}

We perform a proper statistical predictive analysis of temperature peaks exceeding high thresholds, as outlined in Section \ref{sec:strong_results}. We begin by considering the predictive density of peaks exceeding the threshold ${t_E}=34$ °C, obtained setting $\tau_E=\tau_I=1-k/n$, which implies $\tau^\star=1$ in formula \eqref{eq:GP_predictive_dens}. Frequentist estimates of this predictive density are obtained by substituting the ML and GPWM estimates of $\bftheta$ in $ h_{\bftheta, \tau_E}^\star$ (see Section \ref{sec:freq}). Meanwhile, the Bayesian estimate is derived from the posterior predictive density in \eqref{eq:Bayes_predictive_den}, approximated using the Monte Carlo method
$$
\widehat{h}_{{\tau_E}}^{{\text{(B)}}}(y) \approx \sum_{i=1}^M 
h_{\widetilde{\bftheta}_i}
\left(
(y-X_{n-k,n}){\tau^\star}^{\widetilde{\gamma}_i}-
\widetilde{\sigma}_i \frac{1-{\tau^\star}^{\widetilde{\gamma}_i}}{\widetilde{\gamma}_i}
\right)
{\tau^\star}^{\widetilde{\gamma}_i},
$$
where $\widetilde{\bftheta}_i=(\widetilde{\gamma}_i, \widetilde{\sigma}_i)$, with $i=1,\ldots,M$, represents a sample from the posterior distribution of the GP parameters. Notably, when $\tau^\star=1$ the formula simplifies accordingly. The fourth row of Table \ref{tab:_real_analysis} presents A95\%PI derived from these estimated densities. Since the predictive densities are computed using the intermediate level $\tau_E=\tau_I$, it is unsurprising that all three methods yield the same interval. This finding suggests that, given a temperature above 34°C—an event expected approximately 5.4\% of the time—it is plausible that the observed temperature falls within the range of approximately 34.1°C to 37.6°C.

To account for even more extreme events, we estimate the predictive density of future peaks exceeding higher thresholds. We explain a simple and interpretable criterion for selecting these thresholds.
For a short-tailed distribution $F$, with $\gamma<0$, we known that for any $x>0$, the result 
$$
\frac{x_E-Q(1-xu)}{x_E-Q(1-u)} \to x^{-\gamma},\quad u\to 0
$$
holds, see \cite[][Theorem 1.1.13]{dehaan+f06}. Now, setting $u=1-\tau_I$ with $\tau_I=1-k/n$ and defining $x=\tau^\star$, where $\tau_E\geq \tau_I$ represents an extreme level used to determine a very high threshold, this condition implies
$$
x_E-Q(\tau_E)\approx (\tau^\star)^{-\gamma} (x_E-Q(\tau_I))\quad n\to\infty.
$$
Introducing a scaling factor $c=(\tau^{\star})^{\gamma}$, we obtain the final condition
$$
x_E-Q(\tau_E)\approx \frac{x_E-Q(\tau_I)}{c}, \quad n\to\infty.
$$
This formulation allows us to set $c>0$ such that the gap between the right endpoint $x_E$ of $F$ and the extreme threshold $t_E=Q(\tau_E)$ is, for instance, half (c=2), a third (c=3), or a quarter (c=4) of the gap between $x_E$  and the initial threshold $t_I=Q(\tau_I)$. We specify $c=2,3,4$ and estimate $\tau_E=1 - c^{1/\gamma} (1-\tau_I)$ along with the extreme threshold $Q(\tau_E)$.

The last three rows of Table \ref{tab:_real_analysis} report the estimated values of $\widehat{\tau}_E=c^{1/\widehat{\gamma}} (1-\tau_I)$ in percentage, and $\widehat{Q}(\widehat{\tau}_E)$, according to \eqref{eq:est_extreme_quantile} in °C.
The parameter estimates $(\widehat{\gamma},\widehat{\sigma})$ are obtained using the ML (first and second columns) and GPWM (fourth and fifth columns) methods. The table also presents A95\%CI for $\tau_E$ and $Q(\tau_E)$ (seventh and eighth columns), derived from their empirical posterior obtained using the formulas $\widetilde{\tau}_E^{(i)}=c^{1/\widetilde{\gamma}_i} (1-\tau_I)$ and 
 $$
 \widetilde{Q}_i(\widetilde{\tau}_E^{(i)})=X_{n-k,n}+\widetilde{\sigma}
\frac{\left(\frac{1-\widetilde{\tau}_E^{(i)}}{1-\tau_I}\right)^{-\widetilde{\gamma}_i} -1}{\widetilde{\gamma}_i},
$$
where $(\widetilde{\gamma}_i,\widetilde{\sigma}_i)$, with $i=1,\ldots,M$, is a posterior sample of the GP parameters.
The thresholds estimated using the GPWM method are consistently higher than those obtained with the ML approach, with the gap widening as $c$ increases. In contrast, the Bayesian approach suggests a broader range of plausible values.
The bottom panels of Figure \ref{fig:real_analysis} show the estimated predictive densities obtained using the ML (yellow dotted lines), GPWM (violet dashed lines), and Bayesian (green solid lines) approaches. As expected, the divergence between the predictive densities increases with higher values of $c$. The GPWM method assigns progressively more mass to hotter temperatures compared to the ML method. In contrast, the Bayesian approach appears more balanced, yielding wider predictive densities that account for both lower and higher temperatures.
Table \ref{tab:_real_analysis} (last three rows) presents the corresponding A95\%PI estimated with the ML (3rd column), GPWM (6th column), and Bayesian (9th column) methods. The Bayesian approach suggests that if a temperature exceeds a threshold between 36.2 °C (36.9 °C or 37.2 °C) and 38.2 °C (39.8 °C or 40.5 °C), whose  occurrence probability of such an event ranges between 0.075\% (0.006\% or 0.001\%) and 1.027\% (0.390\% or 0.196\%), then, under this scenario, the expected temperature falls within the range of 36.4 °C (37.0 °C or 37.4 °C) to 39.1 °C (40.2 °C or 41.6 °C), encompassing both the hottest recorded temperature and even higher extremes. The results obtained with the other two estimation methods can be interpreted similarly.

We complete the analysis providing point forecasts of extreme temperatures for the five years following 2023. Specifically, as outlined in Section \ref{sec:tail_risk_assessment}, these forecasts are derived from predictive return levels corresponding to return periods $T=37,\ldots,365\times 5$, whcih are computed as  $(\widehat{H}_{\tau_I}^{\text{(M)}})^{\leftarrow}(1-\tau^\star)$, where $\tau^\star = (1-\tau_E)/(1-\tau_I)$ with $\tau_E=1-1/T$.
The bottom-right panel of Figure \ref{fig:real_analysis} presents these estimated return levels using the ML (violet solid line), PWM (orange solid line), and Bayesian (green solid line) methods. Notably, on August 12, 2024, Milan recorded a peak temperature of 36 °C (indicated by the red dot-dash line). All three inferential approaches predict a return level of approximately 36.8 °C for that date.
To quantify the uncertainty in future high-temperature predictions, we compute A95\% PI over time based on the predictive distribution of peak temperatures at extreme levels $\tau_E = 1-1/T$ and intermediate levels $\tau_I = 1- \tilde{k}/n$, where $\tilde{k}=4n/T$. This ensures that the ratio $\tau^\star=1/4$, meaning the intermediate level is set so that the sample fraction $1-\tau_I$  is four times the exceedance probability $1-\tau_E$.
The dotted lines, overlaid on the point forecasts, represent these predictive intervals for the ML (violet), PWM (orange), and Bayesian (green) methods. Among them, the ML approach provides the most conservative estimates, suggesting, for example, that on August 12, 2024, the peak temperature is likely to fall between 35.6 °C and 37.9 °C with 95\% probability. In contrast, the Bayesian method yields the widest range, with an A95\%PI between 34.6 °C and 38.9 °C.

\subsection{Dynamic prediction of stock market indexes}\label{sec:financial_data}
\begin{figure}[t!]
	\centering
	\includegraphics[width=0.27\textwidth, page=1]{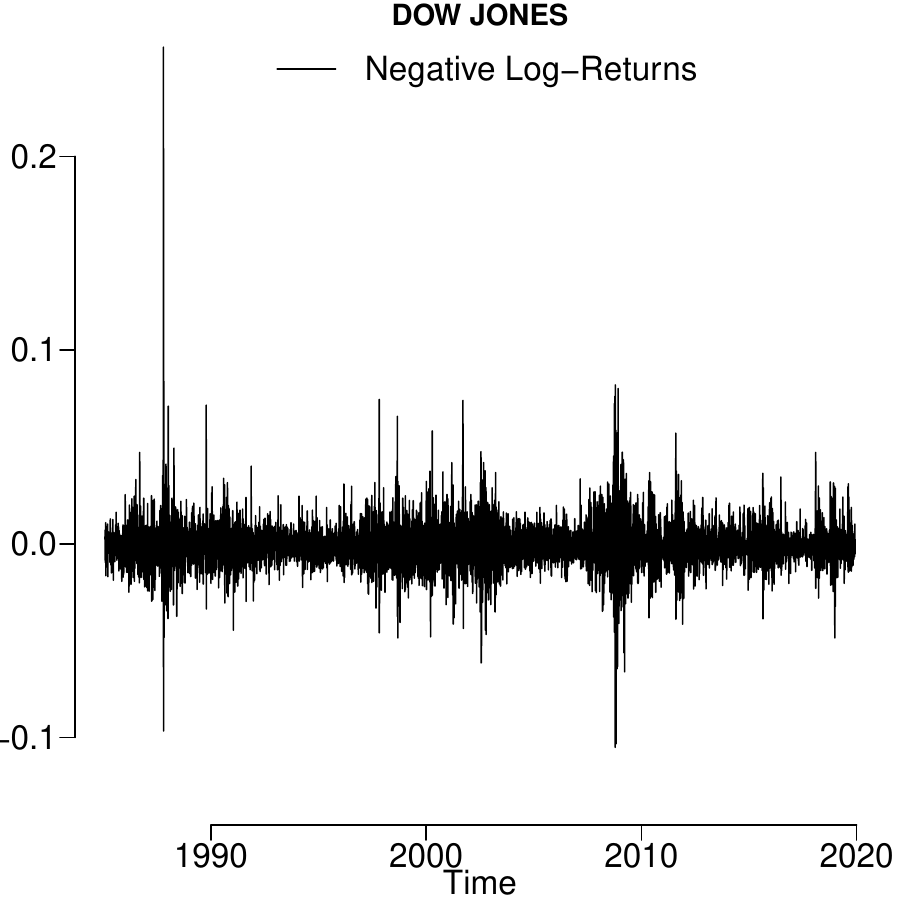}
	\includegraphics[width=0.27\textwidth, page=3]{dowjones}
	\includegraphics[width=0.27\textwidth, page=4]{dowjones}\\
	\includegraphics[width=0.27\textwidth, page=5]{dowjones}
	\includegraphics[width=0.27\textwidth, page=10]{dowjones}
	\includegraphics[width=0.27\textwidth, page=13]{dowjones}
	\includegraphics[width=0.27\textwidth, page=14]{dowjones}
	\includegraphics[width=0.27\textwidth, page=16]{dowjones}
	\includegraphics[width=0.27\textwidth, page=17]{dowjones}
	\caption{The top panels show the Dow Jones log-returns, autocorrelation of squared returns, and their empirical distribution. The middle panels display estimated $\gamma$ values across sample sizes $k$ using Hill, ML, and PWM methods; autocorrelation of squared residuals; and $\gamma$ estimates based on residuals via the Hill method. The bottom-left and center panels report point forecasts of $\widehat{Q}_n^{\text{(M)}}(0.999)$ using rolling windows of 1,000 log-returns, from ML (black) and Bayesian (magenta) methods, with A99\%PI (green and cyan dotted lines). The bottom-right panel shows the predictive density for the unobserved log-return on December 13, 2019, based on the latest 1,000 observations.}\label{fig:downjones_dynamic}
\end{figure}

We analyze the daily negative log-returns of the Dow Jones Industrial Average index from January 29, 1985, to December 12, 2019, comprising a total of $n=8,785$ observations. The top-left panel of Figure \ref{fig:downjones_dynamic} displays the log-return series, which exhibits well-known stylized facts such as heteroskedasticity and heavy tails \citep{embrechts2013modelling}. These characteristics are further supported by the autocorrelation of squared log-returns and the empirical distribution of log-returns, shown in the top-center and top-right panels, respectively. Next, we estimate the extreme value index $\gamma$ using the Hill, ML, and PWM methods \citep[][Ch. 3]{dehaan+f06}. The results, displayed in the middle-left panel of Figure \ref{fig:downjones_dynamic}, suggest that $\gamma$ is likely positive and remains stable when the effective sample size ranges between $100$ and $350$. Within this interval, we estimate $\gamma\approx 0.35$, providing evidence that the marginal distribution is heavy-tailed, albeit with finite first and second moments.

Next, we proceed with the dynamic prediction of future tail events based on past observations. Specifically, we fit various ARMA-GARCH models using a Student-$t$ likelihood on rolling windows of log-returns $(Y_j,\ldots,Y_{j+T-1})$, with $j=1,T,2T\ldots$ and time window $T=1000$, spanning from January 29, 1985, to December 12, 2019.
Using standard model selection criteria such as AIC and BIC, we determine that a GARCH(1,1) model best describes the log-return dynamics. We further validate the model by examining the correlograms of residuals and their squared values (see the middle-center panel of Figure \ref{fig:downjones_dynamic}) and performing weighted Ljung–Box and Li–Mak tests \citep{fisgal2012}, both of which suggest that the residuals can be considered essentially independent. The computations are carried out using the {\tt R} package {\tt rugarch} \citep{rugarch}.
The middle-right panel of Figure \ref{fig:downjones_dynamic} displays the estimates of $\gamma$ for different values of the effective sample size $k$, computed using the Hill estimator along with pointwise 95\% confidence intervals. These estimates are based on residuals from a sample spanning December 15, 2015, to December 12, 2019. The results indicate that $\widehat{\gamma}_n$ remains stable for effective sample sizes between 25 and 150, within which we estimate $\gamma\approx 1$. This suggests that the residuals follow a heavy-tailed distribution, and for subsequent computations, we set $k=100$.

The log-returns $\bfY_{T,j}=(Y_j,\ldots,Y_{j+T-1})$, with $j=1,T, 2T, \ldots$ and time window $T=1000$, with observed value $\bfy_{T,j}=(y_j,\ldots,y_{j+T-1})$, are filtered using the estimated GARCH(1,1) model, yielding the residuals $\widehat{\bfvarepsilon}_{T,j}=(\widehat{\varepsilon}_{j}^{(T)},\ldots, \widehat{\varepsilon}_{j+T-1}^{(T)})$. These residuals serve as the basis for predicting future tail events by analyzing the ``out-of-sample'' residuals $\widehat{\varepsilon}_{n+1}^{(n)}=\widehat{\varepsilon}_{j+T}^{(T)}$,  {with} $j=1,T, 2T, \ldots$, as outlined in Section \ref{sec:time_series}. Specifically, using the residuals, we apply the frequentist and Bayesian estimators introduced in Sections \ref{sec:freq} and \ref{sec:bayes} to estimate the predictive distribution $\Prob(\varepsilon_{j+T}\leq z \mid \varepsilon_{j+T}>Q(\tau_E), \bfvarepsilon_{T,j})$, where $ \bfvarepsilon_{T,j}=(\varepsilon_{j},\ldots, \varepsilon_{j+T-1})$. For each estimated predictive distribution obtained for $j=1,T, 2T, \ldots$, we compute using ML, PWM and the Bayesian methods the point forecasts $\widehat{Q}_n^{\text{(M)}}(\tau_E)$, with $\tau_E=0.999$ (see Section \ref{sec:tail_risk_assessment}), along with the corresponding 99\% predictive intervals. Then, following the methodology outlined in Section \ref{sec:time_series}, we transform the residual-based predictions back to the original log-returns using the relationship $y=\widehat{\mu}_{j}^{(T)}+ \widehat{\xi}_{j}^{(T)}z$, where fitted values are computed from data $\bfy_{T,j}$. This allows us to obtain estimators for the predictive distribution $\Prob(Y_{j+T}\leq y \mid  Y_{j+T}> Q_{j+T}(\tau_E),  \boldsymbol{S}_{j+T-1})$ which characterizes the peaks of the original log-returns conditional on previously observed values. We recall that, as in Section \ref{sec:time_series}, $Q_{j+T}$ represents the extreme quantile of the conditional distribution of $\bfY_{j+T}$ given past observations.

For brevity, we report only the results obtained with the ML method (solid black line) in the bottom-left panel of Figure \ref{fig:downjones_dynamic}, as the results from the PWM method are very similar. The bottom-center panel displays the point forecasts obtained using the Bayesian approach. Both panels also include 99\% predictive intervals (dotted lines) to assess uncertainty. The point forecasts and predictive intervals appear broadly reliable, effectively capturing the volatility bursts in the data. The predictive intervals seem well-calibrated, as they encompass both the point forecasts and all observed data peaks. Finally, the bottom-right panel of Figure \ref{fig:downjones_dynamic} presents the predictive density of the peak log-return for December 13, 2019, obtained using the Bayesian approach and the last $1,000$ log-returns. The blue diamond represents the largest observed log-return, while the green square marks its point forecast, estimated as ${\widehat{Q}^{\text{(B)}}_{
j+T}}(\tau_E)$, with $\tau_E=1-1/1000$. As expected, the point forecast closely aligns with the largest observed log-return. The red dots indicate the lower and upper bounds of the $99\%$ predictive interval, which successfully encompasses both the maximum log-return and its point forecast.

\section{Conclusion}\label{sec:conclusion}
The statistical prediction of future peaks exceeding high thresholds—across various threshold levels representing different degrees of extrapolation beyond observed data—has been thoroughly investigated in this paper from both theoretical and practical perspectives. 
Our proposed frequentist and Bayesian methods for predicting future tail events are mathematically sound, straightforward to implement, and easy to interpret in practical applications. To our knowledge, rigorous theoretical results on the accuracy of predictive density estimators have been limited if not entirely untouched.
 To ensure that the tools discussed are both interpretable in real-world scenarios and analytically tractable, we have primarily focused on the case of independent observations. Nonetheless, we have demonstrated that our methodology can be readily extended to  linear time series, facilitating one-step-ahead tail forecasting.
%

In many applications, data exhibit temporal dependence, it is then crucial to extend our theoretical and methodological results to more general settings involving stationary sequences of serially dependent observations, in order to guarantee accuracy and reliability of predictions. This is a highly ambitious goal, as it requires addressing several challenges. Key open problems include establishing both the local and global asymptotic behavior of the GP likelihood function and developing methods to effectively incorporate serial dependence for making interpretable multi-step-ahead predictions.  Our analysis aims to specify and estimate predictive distributions for accurately forecasting future extreme events that exceed high thresholds, potentially extending into the far tail of the data distribution. This is particularly valuable for conducting \lq\lq what-if" analyses, enabling the evaluation of various hazardous scenarios, including worst-case events. Such evaluations are crucial for the understanding of the potential impacts of exceptional extreme events. By integrating these predictive methods, decision-makers can better prepare for and mitigate the consequences of extreme occurrences. Our approach differs from the existing literature on probabilistic forecasting \cite[e.g.,][]{gneiting2007probabilistic, gneiting14}, which focus on developing mathematical tools to identify the best forecast among different competitors. In line with that literature, we are also interested in integrating our conditional predictive distribution into an unconditional one that facilitates accurate forecasting of both extreme and ordinary events. Our results on forecasters tail equivalence make an encouraging first step in this direction. Additionally, we aim to incorporate simulations from climate or weather models as covariates into our forecasting procedure to enhance prediction accuracy. All these objectives remain important directions for future research.
\section*{Acknowledgements}
Simone Padoan is supported by the Bocconi Institute for Data Science and Analytics (BIDSA), Italy, thanks also to MUR - Prin 2022 - Prot. 20227YZ9JK.

\bibliographystyle{chicago} 
\bibliography{bibliopm_final3}

\begin{thebibliography}{}

\bibitem[\protect\citeauthoryear{Aitchison and Dunsmore}{Aitchison and
  Dunsmore}{1975}]{Aitchison_Dunsmore_1975}
Aitchison, J. and I.~R. Dunsmore (1975).
\newblock {\em Statistical Prediction Analysis}.
\newblock Cambridge University Press.

\bibitem[\protect\citeauthoryear{Artzner, Delbaen, Eber, and Heath}{Artzner
  et~al.}{1999}]{artzner1999coherent}
Artzner, P., F.~Delbaen, J.-M. Eber, and D.~Heath (1999).
\newblock Coherent measures of risk.
\newblock {\em Mathematical Finance\/}~{\em 9\/}(3), 203--228.

\bibitem[\protect\citeauthoryear{Balkema and de~Haan}{Balkema and
  de~Haan}{1974}]{balkema1974residual}
Balkema, A.~A. and L.~de~Haan (1974).
\newblock Residual life time at great age.
\newblock {\em Ann. Probab.\/}~{\em 2\/}(5), 792--804.

\bibitem[\protect\citeauthoryear{Bernardo}{Bernardo}{2011}]{BERNARDO2011263}
Bernardo, J.~M. (2011).
\newblock Modern bayesian inference: Foundations and objective methods.
\newblock In P.~S. Bandyopadhyay and M.~R. Forster (Eds.), {\em Philosophy of
  Statistics}, Volume~7 of {\em Handbook of the Philosophy of Science}, pp.\
  263--306. North-Holland.

\bibitem[\protect\citeauthoryear{Bjarnadottir, Li, and Stewart}{Bjarnadottir
  et~al.}{2019}]{BJARNADOTTIR2019271}
Bjarnadottir, S., Y.~Li, and M.~G. Stewart (2019).
\newblock {C}limate {A}daptation for {H}ousing in {H}urricane {R}egions.
\newblock In E.~Bastidas-Arteaga and M.~G. Stewar (Eds.), {\em Climate
  Adaptation Engineering}, pp.\  271--299. Butterworth-Heinemann.

\bibitem[\protect\citeauthoryear{B{\"u}cher and Segers}{B{\"u}cher and
  Segers}{2017}]{b+s2017}
B{\"u}cher, A. and J.~Segers (2017).
\newblock {On the maximum likelihood estimator for the Generalized
  Extreme-Value distribution}.
\newblock {\em Extremes\/}~{\em 20}, 839--872.

\bibitem[\protect\citeauthoryear{Butler}{Butler}{1986}]{butler1986predictive}
Butler, R.~W. (1986).
\newblock Predictive likelihood inference with applications.
\newblock {\em Journal of the Royal Statistical Society Series B: Statistical
  Methodology\/}~{\em 48\/}(1), 1--23.

\bibitem[\protect\citeauthoryear{Coles}{Coles}{2001}]{coles2001}
Coles, S. (2001).
\newblock {\em An introduction to statistical modeling of extreme values}.
\newblock Springer.

\bibitem[\protect\citeauthoryear{Coles and Pericchi}{Coles and
  Pericchi}{2003}]{coles2003}
Coles, S. and L.~Pericchi (2003).
\newblock Anticipating catastrophes through extreme value modelling.
\newblock {\em Journal of the Royal Statistical Society: Series C (Applied
  Statistics)\/}~{\em 52}, 405--416.

\bibitem[\protect\citeauthoryear{Coles and Powell}{Coles and
  Powell}{1996}]{coles1996bayesian}
Coles, S.~G. and E.~A. Powell (1996).
\newblock Bayesian methods in extreme value modelling: a review and new
  developments.
\newblock {\em International Statistical Review/Revue Internationale de
  Statistique\/}~{\em 64}, 119--136.

\bibitem[\protect\citeauthoryear{Davison}{Davison}{1986}]{davison1986approximate}
Davison, A.~C. (1986).
\newblock Approximate predictive likelihood.
\newblock {\em Biometrika\/}~{\em 73\/}(2), 323--332.

\bibitem[\protect\citeauthoryear{Davison and Smith}{Davison and
  Smith}{1990}]{davison1990}
Davison, A.~C. and R.~L. Smith (1990).
\newblock Models for exceedances over high thresholds.
\newblock {\em Journal of the Royal Statistical Society Series B Statistical
  Methodology\/}~{\em 52\/}(3), 393--425.

\bibitem[\protect\citeauthoryear{Dawid}{Dawid}{1984}]{dawid1984present}
Dawid, A.~P. (1984).
\newblock Present position and potential developments: Some personal views
  statistical theory the prequential approach.
\newblock {\em Journal of the Royal Statistical Society: Series A
  (General)\/}~{\em 147\/}(2), 278--290.

\bibitem[\protect\citeauthoryear{de~Carvalho}{de~Carvalho}{2016}]{de2016statistics}
de~Carvalho, M. (2016).
\newblock Statistics of extremes: Challenges and opportunities.
\newblock {\em Extreme events in finance: A handbook of extreme value theory
  and its applications\/}, 195--213.

\bibitem[\protect\citeauthoryear{de~Haan and Ferreira}{de~Haan and
  Ferreira}{2006}]{dehaan+f06}
de~Haan, L. and A.~Ferreira (2006).
\newblock {\em Extreme Value Theory: An Introduction}.
\newblock Springer.

\bibitem[\protect\citeauthoryear{de~Haan and Resnick}{de~Haan and
  Resnick}{1996}]{deHaan96}
de~Haan, L. and S.~Resnick (1996).
\newblock Second-order regular variation and rates of convergence in
  extreme-value theory.
\newblock {\em The Annals of Probability\/}~{\em 24\/}(1), 97--124.

\bibitem[\protect\citeauthoryear{Dombry and Ferreira}{Dombry and
  Ferreira}{2019}]{dombry19}
Dombry, C. and A.~Ferreira (2019).
\newblock {Maximum likelihood estimators based on the block maxima method}.
\newblock {\em Bernoulli\/}~{\em 25\/}(3), 1690--1723.

\bibitem[\protect\citeauthoryear{Dombry, Padoan, and Rizzelli}{Dombry
  et~al.}{2025}]{dombry23}
Dombry, C., S.~A. Padoan, and S.~Rizzelli (2025).
\newblock {Asymptotic theory for Bayesian inference and prediction: from the
  ordinary to a conditional Peaks-Over-Threshold method}.
\newblock {\em arXiv preprint arXiv:2310.06720v2\/}.

\bibitem[\protect\citeauthoryear{Drees, Ferreira, and de~Haan}{Drees
  et~al.}{2004}]{drees04}
Drees, H., A.~Ferreira, and L.~de~Haan (2004).
\newblock On maximum likelihood estimation of the extreme value index.
\newblock {\em Annals of Applied Probability\/}~{\em 14\/}(1), 1179--1201.

\bibitem[\protect\citeauthoryear{Embrechts, Kl{\"u}ppelberg, and
  Mikosch}{Embrechts et~al.}{2013}]{embrechts2013modelling}
Embrechts, P., C.~Kl{\"u}ppelberg, and T.~Mikosch (2013).
\newblock {\em Modelling extremal events: for insurance and finance},
  Volume~33.
\newblock Springer Science \& Business Media.

\bibitem[\protect\citeauthoryear{Emmer, Kratz, and Tasche}{Emmer
  et~al.}{2015}]{marie2015}
Emmer, S., M.~Kratz, and D.~Tasche (2015).
\newblock What is the best risk measure in practice? a comparison of standard
  measures.
\newblock {\em Journal of Risk\/}~{\em 18\/}(2), 31--60.

\bibitem[\protect\citeauthoryear{Falk, H{\"u}sler, and Reiss}{Falk
  et~al.}{2010}]{falk2010}
Falk, M., J.~H{\"u}sler, and R.-D. Reiss (2010).
\newblock {\em Laws of Small Numbers: Extremes and Rare Events}.
\newblock Springer.

\bibitem[\protect\citeauthoryear{Fisher and Gallagher}{Fisher and
  Gallagher}{2012}]{fisgal2012}
Fisher, T.~J. and C.~M. Gallagher (2012).
\newblock New weighted portmanteau statistics for time series goodness of fit
  testing.
\newblock {\em Journal of the American Statistical Association\/}~{\em 107},
  777--787.

\bibitem[\protect\citeauthoryear{Galanos and Kley}{Galanos and
  Kley}{2022}]{rugarch}
Galanos, A. and T.~Kley (2022).
\newblock {\em rugarch: Univariate {GARCH} Models}.
\newblock R package v. 1.4--6.

\bibitem[\protect\citeauthoryear{Geisser}{Geisser}{1993}]{geisser1993predictive}
Geisser, S. (1993).
\newblock {\em Predictive Inference}, Volume~55.
\newblock CRC Press.

\bibitem[\protect\citeauthoryear{Girard, Stupfler, and Usseglio-Carleve}{Girard
  et~al.}{2021}]{g+s+u2021}
Girard, S., G.~Stupfler, and A.~Usseglio-Carleve (2021).
\newblock {Extreme conditional expectile estimation in heavy-tailed
  heteroscedastic regression models}.
\newblock {\em The Annals of Statistics\/}~{\em 49}, 3358 -- 3382.

\bibitem[\protect\citeauthoryear{Gneiting, Balabdaoui, and Raftery}{Gneiting
  et~al.}{2007}]{gneiting2007probabilistic}
Gneiting, T., F.~Balabdaoui, and A.~E. Raftery (2007).
\newblock Probabilistic forecasts, calibration and sharpness.
\newblock {\em Journal of the Royal Statistical Society Series B: Statistical
  Methodology\/}~{\em 69\/}(2), 243--268.

\bibitem[\protect\citeauthoryear{Gneiting and Katzfuss}{Gneiting and
  Katzfuss}{2014}]{gneiting14}
Gneiting, T. and M.~Katzfuss (2014).
\newblock Probabilistic forecasting.
\newblock {\em Annual Review of Statistics and Its Application\/}~{\em 1},
  125--151.

\bibitem[\protect\citeauthoryear{Gomes, Henriques-Rodrigues, Fraga~Alves, and
  Manjunath}{Gomes et~al.}{2013}]{ivete2013}
Gomes, M.~I., L.~Henriques-Rodrigues, M.~I. Fraga~Alves, and B.~G. Manjunath
  (2013).
\newblock Adaptive port-mvrb estimation: an empirical comparison of two
  heuristic algorithms.
\newblock {\em Journal of Statistical Computation and Simulation\/}~{\em
  83\/}(6), 1129--1144.

\bibitem[\protect\citeauthoryear{Hinkley}{Hinkley}{1979}]{hinkley1979predictive}
Hinkley, D. (1979).
\newblock Predictive likelihood.
\newblock {\em The Annals of Statistics\/}~{\em 7\/}(4), 718--728.

\bibitem[\protect\citeauthoryear{Kratz, Lok, and McNeil}{Kratz
  et~al.}{2018}]{marie2018}
Kratz, M., H.~Y. Lok, and J.~A. McNeil (2018).
\newblock Multinomial var backtests: A simple implicit approach to backtesting
  expected shortfall.
\newblock {\em Journal of Banking \& Finance\/}~{\em 88}, 393--407.

\bibitem[\protect\citeauthoryear{Leadbetter, Lindgren, and
  Rootz{\'e}n}{Leadbetter et~al.}{1983}]{leadbetter1983extremes}
Leadbetter, M.~R., G.~Lindgren, and H.~Rootz{\'e}n (1983).
\newblock {\em Extremes and related properties of random sequences and
  processes}.
\newblock Springer.

\bibitem[\protect\citeauthoryear{Padoan and Rizzelli}{Padoan and
  Rizzelli}{2024a}]{padoan24b}
Padoan, S.~A. and S.~Rizzelli (2024a).
\newblock Empirical bayes inference for the block maxima method.
\newblock {\em Bernoulli\/}~{\em 30\/}(3), 2154--2184.

\bibitem[\protect\citeauthoryear{Padoan and Rizzelli}{Padoan and
  Rizzelli}{2024b}]{padoan24}
Padoan, S.~A. and S.~Rizzelli (2024b).
\newblock Strong convergence of peaks over a threshold.
\newblock {\em Journal of Applied Probability\/}~{\em 61\/}(2), 529--539.

\bibitem[\protect\citeauthoryear{Raoult and Worms}{Raoult and
  Worms}{2003}]{raoult03}
Raoult, J.-P. and R.~Worms (2003).
\newblock Rate of convergence for the generalized {P}areto approximation of the
  excesses.
\newblock {\em Advances in Applied Probability\/}~{\em 35\/}(4), 1007--1027.

\bibitem[\protect\citeauthoryear{Rootz{\'e}n and Katz}{Rootz{\'e}n and
  Katz}{2013}]{rootzen2013design}
Rootz{\'e}n, H. and R.~W. Katz (2013).
\newblock Design life level: quantifying risk in a changing climate.
\newblock {\em Water Resources Research\/}~{\em 49\/}(9), 5964--5972.

\bibitem[\protect\citeauthoryear{Smith}{Smith}{1997}]{smith1997predictive}
Smith, R. (1997).
\newblock Predictive inference, rare events and hierarchical models.
\newblock {\em Preprint, University of North Carolina\/}.

\bibitem[\protect\citeauthoryear{Smith}{Smith}{1985}]{smith1985maximum}
Smith, R.~L. (1985).
\newblock Maximum likelihood estimation in a class of nonregular cases.
\newblock {\em Biometrika\/}~{\em 72}, 67--90.

\bibitem[\protect\citeauthoryear{Smith}{Smith}{1999}]{smith1999}
Smith, R.~L. (1999).
\newblock Bayesian and frequentist approaches to parametric predictive
  inference.
\newblock In {\em Bayesian Statistics 6: Proceedings of the Sixth Valencia
  International Meeting June 6-10, 1998}. Oxford University Press.

\bibitem[\protect\citeauthoryear{van~der Vaart}{van~der Vaart}{2000}]{vdv2000}
van~der Vaart, A. (2000).
\newblock {\em {Asymptotic Statistics}}.
\newblock Cambridge University Press.

\end{thebibliography}

\end{document}